\documentclass[runningheads]{llncs}

\usepackage{eccv}

\usepackage{eccvabbrv}

\usepackage{graphicx}
\usepackage{booktabs}
\usepackage{multirow}
\usepackage{colortbl}
\definecolor{mygray}{gray}{.75}
\usepackage[accsupp]{axessibility}  

\usepackage[pagebackref,breaklinks,colorlinks,citecolor=eccvblue]{hyperref}

\usepackage{orcidlink}

\begin{document}

\title{Online Video Quality Enhancement with Spatial-Temporal Look-up Tables} 

\titlerunning{Online Video Quality Enhancement with Spatial-Temporal Look-up Tables}




\author{Zefan Qu\inst{1}\orcidlink{0009-0006-5347-6426} \and
Xinyang Jiang\inst{2}\orcidlink{0000-0002-4991-0596} \and
Yifan Yang\inst{2}\orcidlink{0000-0002-5481-2851} \and
Dongsheng Li\inst{2} \and
Cairong Zhao\inst{1}}

\authorrunning{Z.Qu et al.}

\institute{Tongji University, Shanghai, China \and
Microsoft Research Asia, Shanghai, China\\
\email{\{qzf, zhaocairong\}@tongji.edu.cn} \\ \email{\{xinyangjiang,yifanyang,dongsheng.li\}@microsoft.com}}

\maketitle

\begin{abstract}
  Low latency rates are crucial for online video-based applications, such as video conferencing and cloud gaming, which make improving video quality in online scenarios increasingly important. However, existing quality enhancement methods are limited by slow inference speed and the requirement for temporal information contained in future frames, making it challenging to deploy them directly in online tasks. In this paper, we propose a novel method, STLVQE, specifically designed to address the rarely studied online video quality enhancement (Online-VQE) problem.
Our STLVQE designs a new VQE framework which contains a Module-Agnostic Feature Extractor that greatly reduces the redundant computations and redesign the propagation, alignment, and enhancement module of the network. A Spatial-Temporal Look-up Tables (STL) is proposed, which extracts spatial-temporal information in videos while saving substantial inference time. To the best of our knowledge, we are the first to exploit the LUT structure to extract temporal information in video tasks. Extensive experiments on the MFQE 2.0 dataset demonstrate our STLVQE achieves a satisfactory performance-speed trade-off.
  \keywords{Online Video Quality Enhancement \and Look-up Table \and Video Compression Artifact Reduction}
\end{abstract}

\section{Introduction}
\label{sec:intro}

\begin{figure}[t]
  \centering
  \includegraphics[width=\linewidth,trim=0 90 5 90,clip]{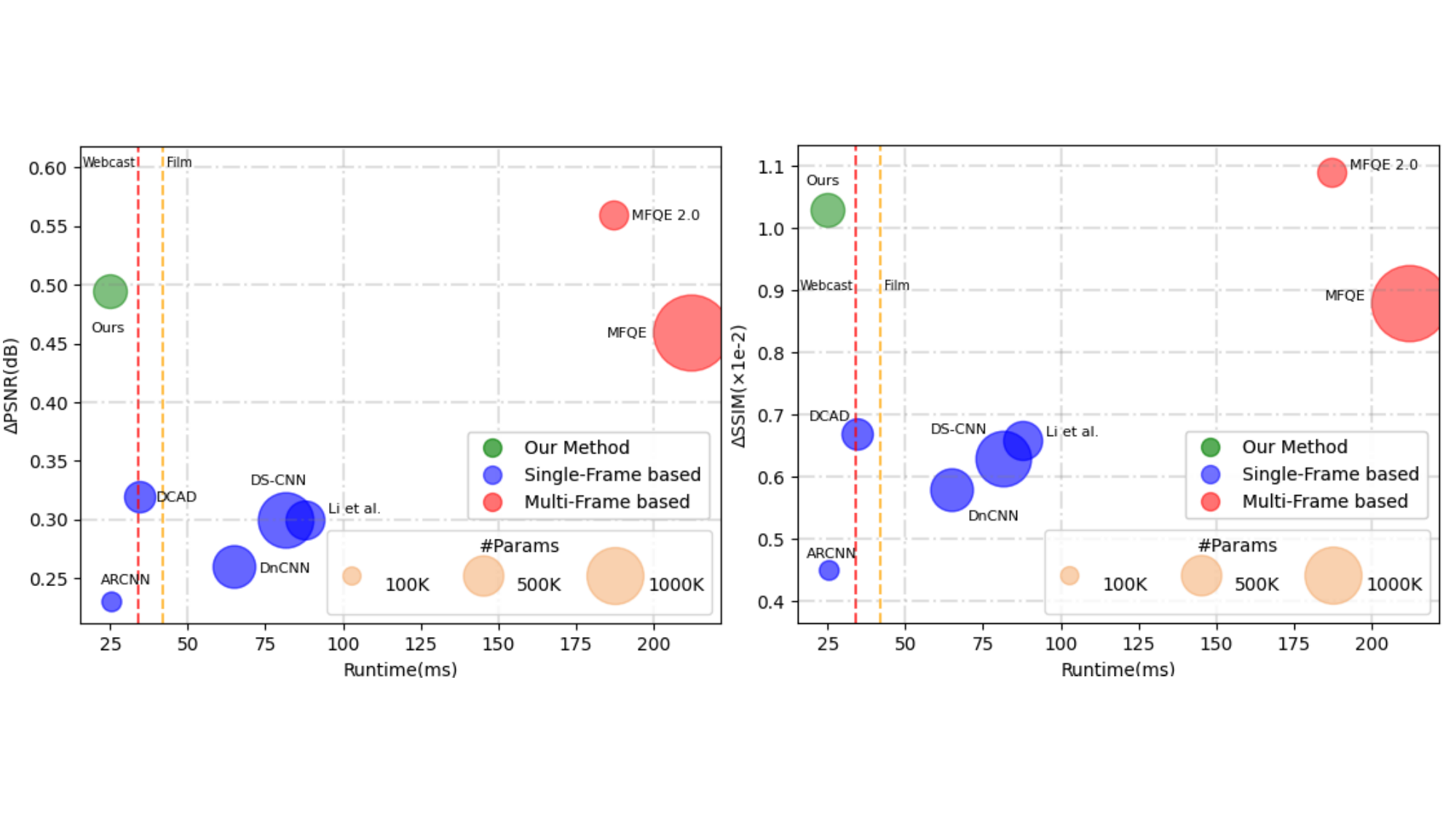}
  \caption{Comparison of $\Delta$PSNR/$\Delta$SSIM, average runtime per frame(720P video)  and parameters on MFQE 2.0 test set at QP=37. Our STLVQE method achieves a great trade-off in enhancement performance and inference speed.}\label{window}
\end{figure}

With the rapid development of social media, users are increasingly seeking high-quality and real-time video content.
In recent times, video-based applications such as video conferencing, webcasting, and cloud gaming have increasingly demanded low latency requirements. This has resulted in higher video compression rates, which can cause a noticeable decrease in video quality and an overall reduction in the Quality of Experience (QoE).
As a result, online Video Quality Enhancement (Online-VQE), which improves the quality of real-time streaming video under online conditions and minimizing video defects caused by compression algorithms (e.g., H.264/AVC~\cite{wiegand2003overview} and H.265/HEVC~\cite{sullivan2012overview}) has become a significant problem. 
However, Online-VQE poses two main challenges compared to the conventional offline VQE approach. 
Firstly, it needs to process high-resolution videos in real-time to ensure a smooth viewing experience.  
Secondly, online video processing techniques~\cite{lin2019tsm,kondratyuk2021movinets,maggioni2021efficient} can only rely on the current and previous frames for inference due to the uncontrolled latency caused by waiting for future frames, which can result in overall video delay.

In recent years, numerous neural network-based techniques for enhancing the quality of compressed video have been developed, resulting in significant advancements in this field. These methods can be broadly classified into two categories, namely single-image based VQE method~\cite{chen2018dpw,jin2018quality,yang2017decoder, yang2018enhancing} and multi-frame based VQE method~\cite{deng2020spatio,guan2019mfqe,lu2018deep,zhao2021recursive,chan2022basicvsr++}. Single-image based approaches are relatively faster in inference speed, but they ignore the significance of temporal information in video~\cite{yang2018multi}. 
On the other hand, multi-frame based VQE achieves state-of-the-art performance by utilizing temporal information. However, most of these approaches are not suitable for online scenarios because they rely on subsequent frames as reference frames and have un-satisfactory inference speeds. Therefore, developing a new online-VQE method that satisfies the aforementioned two constraints while effectively leveraging temporal information is essential.

Reasons that limit the speed and latency of the above temporal VQE methods can be summarized in the following two points: First, a high level of computational redundancy exists within these methods, and these homogeneous computations increase the network's inference time. Second, most of these methods employ complex and computationally intensive models, which sacrificing inference speed and latency for some of the enhancement capabilities. In this work, a more applicable VQE method for online scenarios are proposed by advancing these two perspectives. 

As discussed in previous works~\cite{chan2021basicvsr}, modern VQE methods exploit temporal information by constructing a neural network containing three main components: the propagation, alignment and enhancement module.
In all three modules, similar feature extraction operations on the original image may be involved, e.g., the reference frame selection~\cite{yang2018multi,guan2019mfqe} in propagation module, the optical flow calculation~\cite{chan2021basicvsr, chan2022basicvsr++} in alignment module and the final result obtaining in the enhancement module~\cite{guan2019mfqe, deng2020spatio}.
Most of the state-of-the-art methods adopts distinct feature extractors in different components and obtaining latent feature for each component separately, causing large computation redundancy. 
In this paper, we design a new VQE network framework containing a Module-Agnostic Feature Extractor (MAFE), where the extracted latent feature are uniformly share across three modules. This design allows the network to greatly reduce the redundant feature computation steps and speed up the inference of the network. 
When processing high-resolution videos in VQE task, the complex network of the sota method will greatly increase the computation times of the network, thus affecting the online deployment of the method. Therefore, we in our proposed framework: (1) Redesign the structure of the three modules with a light-weight convolutional structure. 
(2) Propose a Spatial-Temporal Look-up Tables (ST-LUTs) that extend conventional LUTs~\cite{jo2021practical} to conduct queries on both spatial and temporal dimensions,
which allows the LUTs to simultaneously capture information in both temporal and spatial dimensions, facilitating efficient extraction of temporal information. To the best of our knowledge, ST-LUTs is the first approach to employ a LUT structure for the extraction of temporal information.

The major contributions are summarized as follows:
(1) We propose the first online-VQE method STLVQE that achieves real-time process speed, which reduces redundant feature extraction computations while redesigning the propagation, alignment and enhancement modules of the temporal VQE network. 
(2) A Spatial-Temporal Look-up Tables structure is first time proposed for fully exploiting the temporal and spatial information in the video. 
(3) Extensive experiments on MFQE 2.0 dataset demonstrate that our approach achieves a comparable speed-performance trade-off. We also test most of the previous VQE methods on prevalent GPU to provide performance and speed benchmarks, facilitating future research in this area.

\section{Related Works}

\label{sec:formatting}

\textbf{Compressed Video Quality Enhancement.} Recently, numerous works have focused on the degradation of the network video quality due to the lossy video compression algorithms. Existing Compressed VQE work can be broadly classified into two categories, which are single-image based method~\cite{chen2018dpw,dong2015compression,guo2016building,li2017efficient,zhang2017beyond,dai2017convolutional,jin2018quality,yang2017decoder} and multi-frame based method~\cite{deng2020spatio,ding2021patch,guan2019mfqe,yang2018multi,lu2018deep,xu2019non,zhao2021recursive,chan2022basicvsr++}. 


For the single-image based methods, some of the early work~\cite{dong2015compression,li2017efficient,zhang2017beyond,chen2018dpw,guo2016building, yang2018enhancing} are mainly used to enhance the quality of JPEG compressed images, which can also be applied directly to compressed videos to process each frame independently. \cite{dai2017convolutional, jin2018quality, yang2017decoder} utilize certain characteristics of the video compression algorithm or compression parameters to improve the video quality. For example, DS-CNN~\cite{yang2017decoder} and QE-CNN~\cite{yang2018enhancing} use distinct models to enhance the video processed by different compression coding. 

For the multi-frame based methods, MFQE 1.0~\cite{yang2018multi} and MFQE 2.0~\cite{guan2019mfqe} were proposed to use two adjacent high quality frames(PQF) and compute the optical flow between them to enhance the current frame. STDF~\cite{deng2020spatio} uses deformation convolution for an implicit alignment. RFDA~\cite{zhao2021recursive} extracts long-range temporal information from the video via the recursive fusion module. BasicVSR~\cite{chan2021basicvsr} and BasicVSR++~\cite{chan2022basicvsr++} utilize a bidirectional propagation network and a new alignment scheme is proposed in BasicVSR++. All these methods demonstrate the significance of temporal information for solving the compressed VQE problem. 

As far as we know, there is still no compression VQE method for online scenario that has been proposed so far. The above-mentioned methods cannot be directly applied to webcasting environments due to the limitations of the two online  restriction mentioned above.

\textbf{Look-up Table.} Look-up Tables (LUTs) have proven to be highly effective in reducing computational power and time consumption during the inference phase of algorithms by replacing complex computation with fast storage space retrieval operations.
However, the use of LUTs directly is not always flexible as it requires manual adjustment by experts to accurately process images. To overcome this limitation, recent attempts have been made to improve LUTs. ~\cite{zeng2020learning,wang2021real} combine multiple LUTs and learn the weights through neural networks to summing up these LUT results. SR-LUT~\cite{jo2021practical} replaces a small receptive field neural network with a 4D-LUT and achieves good speed-performance trade-offs on SR tasks. SP-LUT~\cite{ma2022learning} and MuLUT~\cite{li2022mulut} makes improvements to the SR-LUT by expanding its receptive field through parallel and serial multiple LUT tables. In this work, we utilize LUTs to extract information in both temporal and spatial dimensions, bringing its advantages to the online VQE task.

\section{Method}

\subsection{Overview}
Given a stream of compressed frames without an explicit ending $\{x_0,...x_i,...\}$, where $x_i \in \mathbb{R}^{C\times H \times W}$ with input channel $C$, height $H$ and width $W$, we sequentially enhance each compressed frame to obtain the target sequence $\{y_0,...y_i,...\}$. Due to the low latency requirement of Online-VQE, obtaining information from future frames to enhance $x_i$ is unacceptable. We only use past and current frames $\{x_0,x_1,...x_i\}$ to predict $y_i$, reducing the video's delay time. Additionally, to avoid stuttering issues, each frame's inference speed must be fast enough.

\begin{figure*}[t]
  \centering
  \includegraphics[width=\linewidth,trim=80 170 100 120,clip]{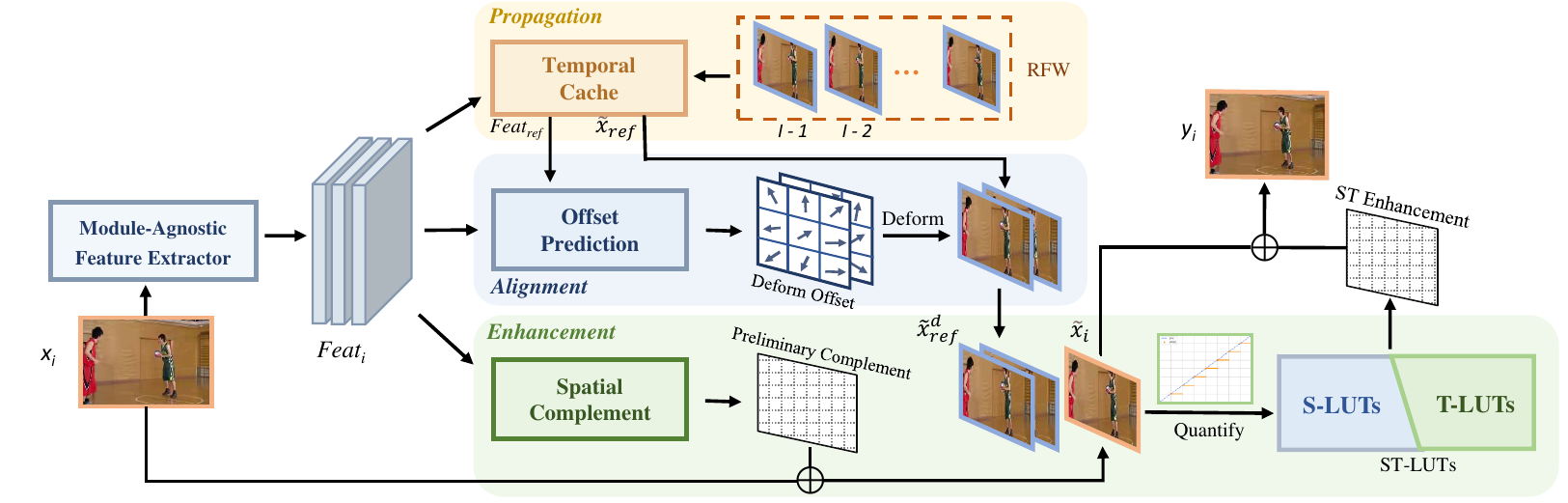}
  \caption{The framework of STLVQE (inference phase), which consists of a Module-Agnostic Feature Extractor and three main parts: the propagation, alignment and enhancement module. In the inference phase, the propagation module selects the reference frame and accesses the relevant information, which is sent to the alignment module for temporal alignment, and finally the aligned reference frames and enhancing frame are input to the enhancement module for the final results. }\label{fig2}
\end{figure*}

The overall flow of our proposed network during inference period is illustrated in Fig.~\ref{fig2} (The difference between the training and the inference phase is illustrated in Subsec. \ref{STLUT} and the figure in appendix). First, a Module-Agnostic Feature Extractor (MAFE) is utilized to extract the feature $Feat_i$ of the current enhancing frame $x_i$ as follows,
\begin{equation}\label{Eq1}
\begin{aligned}
Feat_{i} &= MAFE(x_i),\\
\end{aligned}
\end{equation}
where the $Feat_i$ will be shared by the three subsequent modules, thus reducing the multiple time-consuming extraction steps and fully exploiting the latent frame feature. 

Specifically, in the propagation module, $Feat_i$ is stored in a structure called Temporal Cache, which will provide the temporal information in subsequent modules based on the selection of the reference frames $x_{ref} = \{x_{ref1}, x_{ref2}\}$ without external computation. In the alignment module, $Feat_i$ is concatenate with $Feat_{ref}$, the reference frame feature obtained from the propagation module, to deform $x_{ref}$ with a lightweight offset predictor. In the enhancement module, $Feat_i$ is used to perform a spatial complement, which fully exploits the extracted features and augments the ST-LUTs by increasing its receptive fields.

Under our proposed new framework, only once feature extraction computation for a single frame is required for enhancing each frame.  The extracted features are simultaneously computed with reference frame alignment and preliminary compensation in the current frame space.

\subsection{The Propagation Module}\label{RFW}
The propagation module in the VQE task is responsible for extracting and transmitting temporal information within the video.
Previous methods have explored various approaches, including designing additional structures~\cite{guan2019mfqe} and increasing the number of selected reference frames~\cite{deng2020spatio, zhao2021recursive} to improve network performance, which greatly increase inference time and computational costs of the model.
The architecture of the propagation module is illustrated in Fig.~\ref{window}. The compression configuration parameter quantization parameter (QP) value is utilized as an indicator to select fewer but higher quality frames among multiple candidate frames without extra computation. 
Specifically, a Reference Frame Window (RFW) which contains $W$ frames $\{x_{i-W}.... . x_{i-1}\}$ is set. Then, two frames with the smallest QP values are selected as $x_{ref}$ for the subsequent module.

To mitigate the issue of additional feature extraction operations in previous works, we introduce a cache-like structure known as the Temporal Cache (TC). The TC stores reusable features of past frames that were already extracted by the MAFE in previous process, thereby eliminating the need to re-extract features for each subsequent frame.
In particular, when enhancing the i-th frame $x_i$, STLVQE caches the intermediate feature map $Feat_i$ extracted by the alignment module, as well as the enhanced frame, in the information cache (TC).  
During reference frame selection, the features and enhanced result of $x_{ref1}$ and $x_{ref2}$ are directly retrieved from TC without extra feature extraction process. 

\begin{figure}[t]
  \centering
  \includegraphics[width=0.7\linewidth,trim=100 125 160 30,clip]{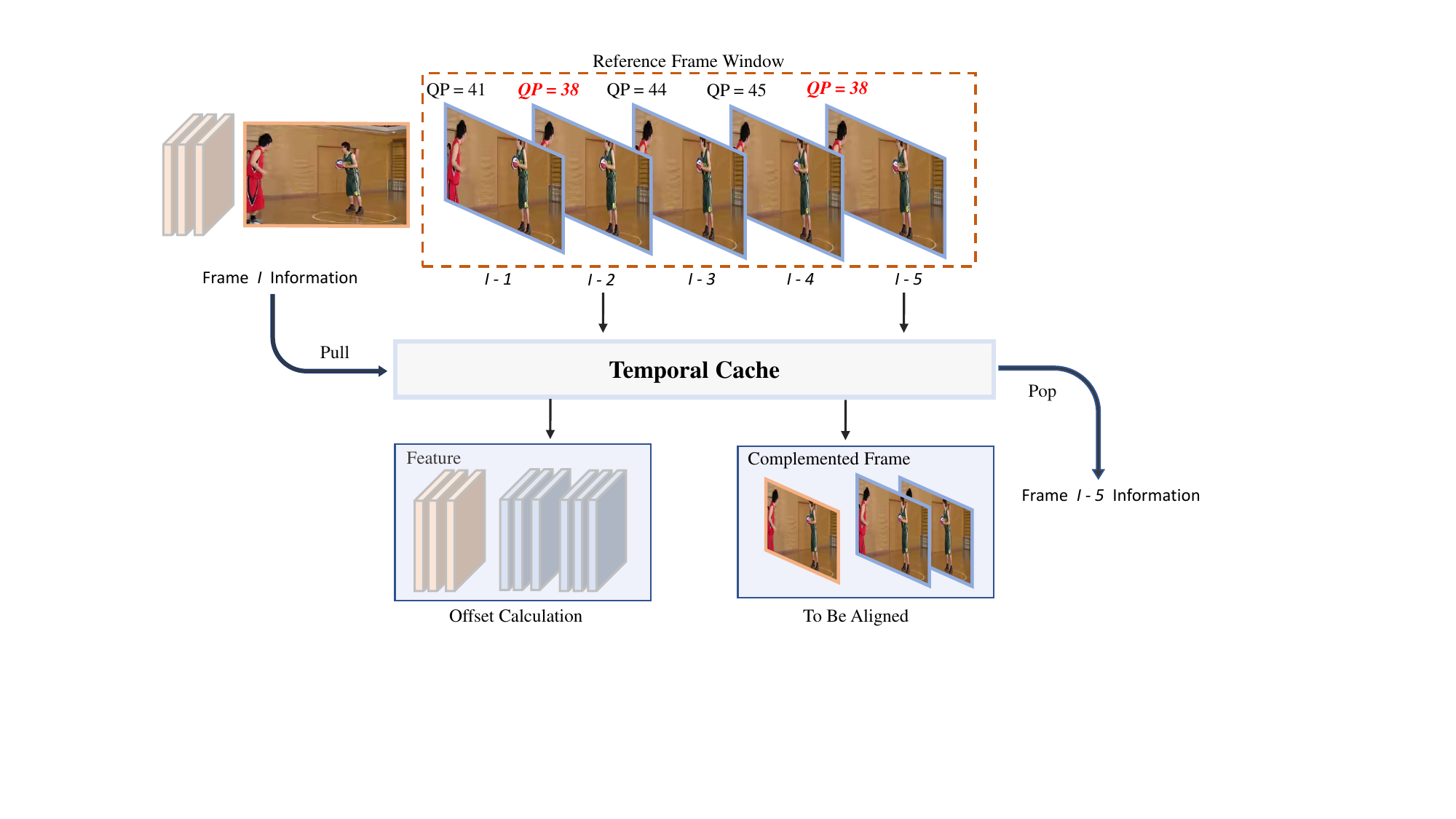}
  \caption{Reference Frame Window and Temporal Cache in the propagation module.}\label{window}
\end{figure}

\subsection{The Alignment Module}\label{alignnetwork}
The alignment module is a crucial component in video processing tasks that handles spatial misalignments between frames, which plays a vital role in effectively utilizing the temporal information embedded in the video, especially when the video content is subject to large motion. STLVQE includes a lightweight offset predictor to replace the optical flow or non-local computation of most VQE methods, which significantly reduce the computation cost. 

The offset predictor is responsible 
for predicting spatial offsets which align correlated pixels in the reference with pixels in $\Tilde{x}_i$.  
As introduced in the last section, the latent features $Feat_{ref}$ and spatially compensated frames $\Tilde{x}_{ref}$ at each processing step are stored in TC, which are retrieved by RFW to enhance future frames. 
Given the feature of a reference frame $\Tilde{x}_{ref}$ retrieved from the TC, as well as the feature of the current frame $\Tilde{x}_i$, the offset predictor predicts $9$ spatial offsets for each pixel in $\Tilde{x}_i$:
\begin{equation}\label{Eq1}
\begin{aligned}
o_{ref} &= C_{offset}([Feat_i, Feat_{ref}^{d}])
\end{aligned}
\end{equation}
where the matrix concatenation is denoted by $[\cdot]$, and  each offset value in $o_{ref}$ represents the position shift of a correlated pixel in $\Tilde{x}_{ref}$ relative to a pixel in $\Tilde{x}_i$. 

The temporal corresponding pixels in the reference frames $x_{ref}$ are spatially aligned with the $3\times3$ pixel offsets $o_{ref}$ and bilinear interpolation.
Each set of obtained correlated pixel values of a pixel in $\Tilde{x}_i$ are re-arranged into a $3\times3$ patch, which are then stacked together to obtain the deformed  reference frame denoted as $\Tilde{x}_{ref}^{d}$ with the size of 
 $N\times C\times3H\times3W$. 



\subsection{The Enhancement Module}\label{STLUT}


The enhancement module contains a spatial compensation network and a ST-LUTs network.
To fully utilize the sharing latent feature $Feat_i$,
a spatial compensation network reuses $Feat_i$ to generate a spatial residual, which compensates the low quality frame $x_i$ to obtain preliminary spatial enhancement before ST-LUTs: 
\begin{equation}\label{Eq1}
\begin{aligned}
\Tilde{x}_i &= x_i + C_{comp}(Feat_i)
\end{aligned}
\end{equation}
where $C_{comp}$ are 3 shallow convolutional layers. 
Since the subsequent enhancement module takes the obtained compensated frame $\Tilde{x}_i$ as input, spatial compensation network can boost the enhancement performance by expanding the receptive field of the ST-LUTs in the enhancement module. To match the resolution of $\Tilde{x}_{ref}^{d}$, $\Tilde{x}_i$ is upsampled by repeating each pixel in $\Tilde{x}_i$ to form a 3x3 patch, then each pixel in the up-sampled $\Tilde{x}_i$ is correlated with the pixel at the same position in $\Tilde{x}_{ref}^{d}$. This is achieved by the unfold and fold operations with different strides as follows:
\begin{equation}\label{upscale}
\begin{aligned}
\Tilde{x}_i^{up} = Fold(Unfold(\Tilde{x}_i, stride = 1), stride = 3),
\end{aligned}
\end{equation}
the up-sampled input frame $\Tilde{x}_i^{up}$ and deformed reference frame $\Tilde{x}_{ref}^{d}$ together serve as the input of the ST-LUTs network for the final quality enhancement. 

Instead of deep neural network, STLVQE adopts a novel Spatial-Temporal Look-up Tables in the enhancement module during inference to replace complex convolutional computation used in training stage with much faster direct memory access, and conducts table queries on both spatial and temporal perspective. 

\textbf{Quantization.} Due to the limited storage space of the LUT, the size of each pixel in the input compensated frame $\Tilde{x}_i^{up}$ and the deformed frames $\Tilde{x}_{ref}^{d}$ needs to be controlled to fit within a suitable range of bits. 
Because the range of values in $\Tilde{x}_i^{up}$ and $\Tilde{x}_{ref}^{d}$ can vary dynamically and the gradient needs to be passed through the quantization operation during end-to-end training, it's not practical to simply quantize the matrices using a fixed quantization hyper-parameter. The enhancement module adopts a Learnable Quantization layer (LSQ) ~\cite{esser2019learned} to quantize the input of the enhancement module as follows:

\begin{equation}\label{Eq1}
\begin{aligned}
\overline{x} &= \lfloor clip(\Tilde{x}/s, -Q_N, Q_P) \rceil, \quad \hat{x} = \overline{x} \times s,
\end{aligned}
\end{equation}
where $clip(z, r_1, r_2)$ returns matrix $z$ with values below $r_1$ set to $r_1$ and values above $r_2$ set to $r_2$, and the $\lfloor z \rceil$ rounds z to the nearest integer. The step size parameter $s$ can be optimized along with the network's training. $\overline{x}$ is the indexes used in the LUTs store and retrieval stages, and $\hat{x}$ is a quantized representation of the data with the same scale as $x$. After quantization, the deep enhancement network and the corresponding LUTs can be constructed with the quantized pixel values as input.


\begin{figure}[t]
  \centering
  \includegraphics[width=0.9\linewidth,trim=20 30 20 5,clip]{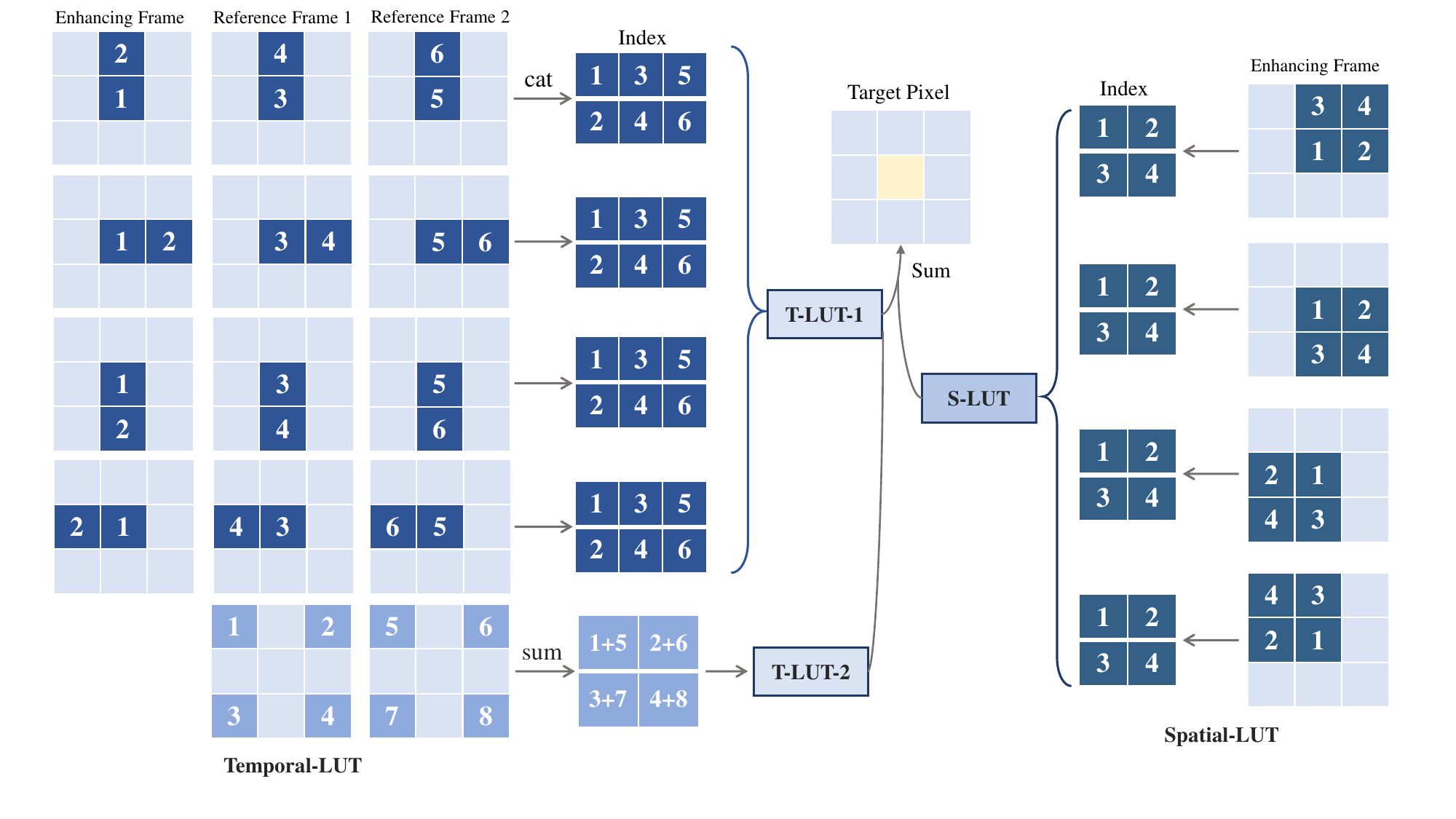}
  \caption{The specific schematic of our Spatial-Temporal Look-up Tables, which consists of two parts: Temporal-LUT (left side) and Spatial-LUT (right side).}\label{LUT}
\end{figure}

\textbf{Deep Enhancement Network Training.} 
In the training phase, a neural network based enhancement module is optimized, which takes the compensated frame $\hat{x}_i^{up}$ and reference frames $\hat{x}_{ref}^{d}$ as input and output the final enhancement results. 
 To train an enhancement network that is suitable for LUT conversion, the receptive field size of the deep network typically cannot exceed 6 (Larger receptive fields will result in a higher storage occupancy, refer to the appendix for details). Therefore, each branch of the enhancement network, which will be replaced by ST-LUTs during the inference stage, consists of 1 convolutional layer with a kernel size of 4 or 6 and 10 convolutional layers with 1$\times$1 kernel. The specific design of the network will be illustrated in the appendix.

\textbf{ST-LUTs Construction \& Replacement.} 
We generate LUTs by permuting all input patches of the receptive field size, and storing all the input/output mappings of the trained deep enhancement network in mapping tables. 
During inference, we retrieve the outputs of the deep enhancement network directly from the LUTs, replacing slow and computationally-intensive convolution  with faster memory access and retain the other components in STLVQE.
Due to the input patch size restriction, conventional LUTs are not suitable for inter-frame queries in the temporal dimension. To address these issues, we firstly propose the ST-LUTs, which employ multiple LUTs to conduct parallel retrieval with different sets of query pixels in both spatial and temporal dimensions.

Specifically, given the quantized compensated frames $\hat{x}_i^{up}$ and the aligned frames $\hat{x}^{d}_{ref}$  as input, the look-up tables are constructed as shown in Fig.~\ref{LUT}, where two sets of LUTs named S-LUTs and T-LUTs are constructed for extracting spatial and temporal information respectively.

For S-LUTs, a rotation ensemble operation~\cite{jo2021practical} is used during the training process to further expand the LUT receptive field. As shown in Fig.~\ref{LUT}, during training, input images are rotated by 0, 90, 180, and 270 degrees, and the results are reverse-rotated and summed up. This approach considers all pixels in a 3$\times$3 receptive field simultaneously. The result of S-LUT $R_i^S$ is expressed as follows:
\begin{equation}\label{upscale}
\begin{aligned}
R_i^S = \frac{1}{4}\sum\limits_{j=0}\limits^{4}{Rot_j^{-1}(f^S(Rot_j(\hat{x}_i^{up})))},
\end{aligned}
\end{equation}
where $Rot_j$ and $Rot_j^{-1}$ are the rotation and reserve-rotation operations, and $f^S$ is the spatial enhancement network or the S-LUTs retrieval. 

For T-LUTs, the pixels at the same location of the enhancing and reference frames will be strongly correlated after alignment, and jointly considering these values can make full use of the temporal information in the video. As shown on the left in Fig.~\ref{LUT}, we establish two T-LUTs: T-LUT-1 take the same position pixels of $\hat{x}_i^{up}$ and $\hat{x}_{ref}^{d}$ as index with the rotation ensemble operation, while T-LUT-2 uses the four edge pixels of each kernel in the $\hat{x}_{ref}^{d}$ as indexes to retrieve the surrounding information of the target pixels. The result of T-LUTs $R_i^T$ can be expressed as follows:
\begin{equation}
\begin{aligned}
R_i^{T1} &= \frac{1}{4}\sum\limits_{j=0}\limits^{4}{Rot_j^{-1}(f^{T1}(Rot_j([\hat{x}_i^{up}, \hat{x}_{ref}^{d}])))},\\
R_i^{T2} &= f^{T2}(\hat{x}_{ref}^{d}),  \quad R_i^{T} = R_i^{T1} + R_i^{T2},
\end{aligned}
\end{equation}
where $f^{T1}$ and $f^{T2}$ are the temporal enhancement network or the T-LUTs retrieval. Then we can acquire the results of ST-LUTs and get the final enhanced image $y_i$ as follows:
\begin{equation}\label{upscale}
\begin{aligned}
R_i &= R_i^S + R_i^T, \quad y_i = \Tilde{x}_i + R_i.
\end{aligned}
\end{equation}

\subsection{Training Scheme}\label{training scheme}
We perform the training of the STLVQE in two stages with an end-to-end fashion. Firstly, the Charbonnier Loss~\cite{charbonnier1994two} is utilized to optimize the model as follows:
\begin{equation}\label{upscale}
\begin{aligned}
L_{char} = \sqrt{(y_t - \overline{y_{t}})^2 + \epsilon},
\end{aligned}
\end{equation}
where $y_t$ and $\overline{y_{t}}$ is the enhanced and ground truth frame at time $t$, and $\epsilon$ is set to $10^{-6}$ in our paper. Secondly, to further enhance the enhancement capability, we perform fine-tuning using L2 loss with a smaller learning rate as follows:
\begin{equation}\label{upscale}
\begin{aligned}
L_{MSE} = \left \|(y_t - \overline{y_{t}})^2\right \|_2^2.
\end{aligned}
\end{equation}

\section{Experiments}
\subsection{Datasets}
Following~\cite{guan2019mfqe,deng2020spatio,zhao2021recursive}, we conduct our experiments on the MFQE 2.0 dataset. It contains totally 126 sequences videos with multi resolutions: SIF(352 $\times$ 240), CIF(352 $\times$ 288), NTSC(720 $\times$ 486), 4CIF(704 $\times$ 576), 240P(416 $\times$ 240), 360P(640 $\times$ 360), 480P(832 $\times$ 280), 720P(1280 $\times$ 720), 1080P(1920 $\times$ 1080) and WQXGA (2560 $\times$ 1600). Same as ~\cite{guan2019mfqe}, the 108 videos in the dataset are used as the training set and 18 videos are used as the test set. All videos were compressed by HM16.5 software in LDP mode with QP values of 22, 27, 32, 37 and 42.

\begin{table*}[t]
\centering
\caption{Overall comparison for $\Delta$ PSNR(dB), $\Delta$ SSIM($\times10^{-2}$), Latency(ms) and FPS over test sequences of MFQE 2.0 Dataset at QP=37. All methods were retested on a NVIDIA RTX 3090 in frame-to-frame manner. Video resolution: Class A(2560$\times$1600), Class B(1980$\times$1080), Class C(832$\times$480), Class D(416$\times$240), Class E(1280$\times$720). }\label{tab:QP37}
\resizebox{\columnwidth}{!}{%
\begin{tabular}{cll|cccc|ccccccccccl}
\hline
\multicolumn{3}{c|}{\multirow{2}{*}{Approach}} & \multicolumn{4}{ c|}{High-Latency VQE Methods} & \multicolumn{11}{c}{Low-Latency VQE Methods} \\ \cline{4-18} 
\multicolumn{3}{c|}{} & \multicolumn{2}{ c|}{MFQE 1.0~\cite{yang2018multi}} & \multicolumn{2}{ c|}{PSTQE~\cite{ding2021patch}} & \multicolumn{2}{c|}{Li et al.~\cite{li2017efficient}} & \multicolumn{2}{c|}{DS-CNN~\cite{yang2018enhancing}} & \multicolumn{2}{c|}{DnCNN~\cite{zhang2017beyond}} & \multicolumn{2}{c|}{DCAD~\cite{wang2017novel}} & \multicolumn{3}{c}{STLVQE(Ours)} \\ \hline
\multicolumn{3}{c|}{Metrics} & \multicolumn{1}{ c|}{$\Delta$P/$\Delta$S} & \multicolumn{1}{ c|}{LT/FPS} & \multicolumn{1}{ c|}{$\Delta$P/$\Delta$S} & \multicolumn{1}{ c|}{LT/FPS} & \multicolumn{1}{c|}{$\Delta$P/$\Delta$S} & \multicolumn{1}{c|}{LT/FPS} & \multicolumn{1}{c|}{$\Delta$P/$\Delta$S} & \multicolumn{1}{c|}{LT/FPS} & \multicolumn{1}{c|}{$\Delta$P/$\Delta$S} & \multicolumn{1}{c|}{LT/FPS} & \multicolumn{1}{c|}{$\Delta$P/$\Delta$S} & \multicolumn{1}{c|}{LT/FPS} & \multicolumn{1}{c|}{$\Delta$P/$\Delta$S} & \multicolumn{2}{c}{LT/FPS} \\ \hline
\multicolumn{1}{c|}{\multirow{2}{*}{A}} & \multicolumn{2}{l|}{Traffic} & \multicolumn{1}{ c|}{0.50/0.90} & \multicolumn{1}{c|}{\multirow{2}{*}{\begin{tabular}[c]{@{} c@{}}1120.1\\ 1.0\end{tabular}}} & \multicolumn{1}{ c|}{0.64/1.04} & \multirow{2}{*}{\begin{tabular}[c]{@{} c@{}}5.6K\\ 0.2\end{tabular}} & \multicolumn{1}{c|}{0.29/0.60} & \multicolumn{1}{c|}{\multirow{2}{*}{\begin{tabular}[c]{@{}c@{}}385.5\\ 2.6\end{tabular}}} & \multicolumn{1}{c|}{0.29/0.60} & \multicolumn{1}{c|}{\multirow{2}{*}{\begin{tabular}[c]{@{}c@{}}374.6\\ 2.7\end{tabular}}} & \multicolumn{1}{c|}{0.24/0.57} & \multicolumn{1}{c|}{\multirow{2}{*}{\begin{tabular}[c]{@{}c@{}}294.9\\ 3.4\end{tabular}}} & \multicolumn{1}{c|}{0.31/0.67} & \multicolumn{1}{c|}{\multirow{2}{*}{\begin{tabular}[c]{@{}c@{}}153.7\\ 6.5\end{tabular}}} & \multicolumn{1}{c|}{\textbf{0.41/0.94}} & \multicolumn{2}{c}{\multirow{2}{*}{\begin{tabular}[c]{@{}c@{}}\textbf{96.3}\\ \textbf{10.4}\end{tabular}}} \\
\multicolumn{1}{c|}{} & \multicolumn{2}{l|}{PeopleOnStreet} & \multicolumn{1}{ c|}{0.80/1.37} & \multicolumn{1}{c|}{} & \multicolumn{1}{ c|}{1.08/1.68} &  & \multicolumn{1}{c|}{0.48/0.92} & \multicolumn{1}{c|}{} & \multicolumn{1}{c|}{0.42/0.85} & \multicolumn{1}{c|}{} & \multicolumn{1}{c|}{0.41/0.82} & \multicolumn{1}{c|}{} & \multicolumn{1}{c|}{0.50/0.95} & \multicolumn{1}{c|}{} & \multicolumn{1}{c|}{\textbf{0.68/1.35}} & \multicolumn{2}{c}{} \\ \hline
\multicolumn{1}{c|}{\multirow{5}{*}{B}} & \multicolumn{2}{l|}{Kimono} & \multicolumn{1}{ c|}{0.50/1.13} & \multicolumn{1}{c|}{\multirow{5}{*}{\begin{tabular}[c]{@{} c@{}}628.1\\ 2.0\end{tabular}}} & \multicolumn{1}{ c|}{0.69/1.36} & \multirow{5}{*}{\begin{tabular}[c]{@{} c@{}}2.9K\\ 0.4\end{tabular}} & \multicolumn{1}{c|}{0.28/0.78} & \multicolumn{1}{c|}{\multirow{5}{*}{\begin{tabular}[c]{@{}c@{}}192.8\\ 5.2\end{tabular}}} & \multicolumn{1}{c|}{0.25/0.75} & \multicolumn{1}{c|}{\multirow{5}{*}{\begin{tabular}[c]{@{}c@{}}178.2\\ 5.6\end{tabular}}} & \multicolumn{1}{c|}{0.24/0.75} & \multicolumn{1}{c|}{\multirow{5}{*}{\begin{tabular}[c]{@{}c@{}}145.4\\ 6.9\end{tabular}}} & \multicolumn{1}{c|}{0.28/0.78} & \multicolumn{1}{c|}{\multirow{5}{*}{\begin{tabular}[c]{@{}c@{}}77.5\\ 12.9\end{tabular}}} & \multicolumn{1}{c|}{\textbf{0.38/1.05}} & \multicolumn{2}{c}{\multirow{5}{*}{\begin{tabular}[c]{@{}c@{}}\textbf{48.6}\\ \textbf{20.6}\end{tabular}}} \\
\multicolumn{1}{c|}{} & \multicolumn{2}{l|}{ParkScene} & \multicolumn{1}{ c|}{0.39/1.03} & \multicolumn{1}{c|}{} & \multicolumn{1}{ c|}{0.49/1.21} &  & \multicolumn{1}{c|}{0.15/0.48} & \multicolumn{1}{c|}{} & \multicolumn{1}{c|}{0.15/0.50} & \multicolumn{1}{c|}{} & \multicolumn{1}{c|}{0.14/0.50} & \multicolumn{1}{c|}{} & \multicolumn{1}{c|}{0.16/0.50} & \multicolumn{1}{c|}{} & \multicolumn{1}{c|}{\textbf{0.33/0.80}} & \multicolumn{2}{c}{} \\
\multicolumn{1}{c|}{} & \multicolumn{2}{l|}{Cactus} & \multicolumn{1}{ c|}{0.44/0.88} & \multicolumn{1}{c|}{} & \multicolumn{1}{ c|}{0.62/1.15} &  & \multicolumn{1}{c|}{0.23/0.58} & \multicolumn{1}{c|}{} & \multicolumn{1}{c|}{0.24/0.58} & \multicolumn{1}{c|}{} & \multicolumn{1}{c|}{0.20/0.48} & \multicolumn{1}{c|}{} & \multicolumn{1}{c|}{0.26/0.58} & \multicolumn{1}{c|}{} & \multicolumn{1}{c|}{\textbf{0.43/0.85}} & \multicolumn{2}{c}{} \\
\multicolumn{1}{c|}{} & \multicolumn{2}{l|}{BQTerrace} & \multicolumn{1}{ c|}{0.27/0.48} & \multicolumn{1}{c|}{} & \multicolumn{1}{ c|}{0.50/0.87} &  & \multicolumn{1}{c|}{0.25/0.48} & \multicolumn{1}{c|}{} & \multicolumn{1}{c|}{0.26/0.48} & \multicolumn{1}{c|}{} & \multicolumn{1}{c|}{0.20/0.38} & \multicolumn{1}{c|}{} & \multicolumn{1}{c|}{0.28/0.50} & \multicolumn{1}{c|}{} & \multicolumn{1}{c|}{\textbf{0.42/0.83}} & \multicolumn{2}{c}{} \\
\multicolumn{1}{c|}{} & \multicolumn{2}{l|}{BasketballDrive} & \multicolumn{1}{ c|}{0.41/0.80} & \multicolumn{1}{c|}{} & \multicolumn{1}{ c|}{0.60/1.04} &  & \multicolumn{1}{c|}{0.30/0.68} & \multicolumn{1}{c|}{} & \multicolumn{1}{c|}{0.28/0.65} & \multicolumn{1}{c|}{} & \multicolumn{1}{c|}{0.25/0.58} & \multicolumn{1}{c|}{} & \multicolumn{1}{c|}{0.31/0.68} & \multicolumn{1}{c|}{} & \multicolumn{1}{c|}{\textbf{0.49/1.00}} & \multicolumn{2}{c}{} \\ \hline
\multicolumn{1}{c|}{\multirow{4}{*}{C}} & \multicolumn{2}{l|}{RaceHorses} & \multicolumn{1}{ c|}{0.34/0.55} & \multicolumn{1}{c|}{\multirow{4}{*}{\begin{tabular}[c]{@{} c@{}}229.8\\ 10.4\end{tabular}}} & \multicolumn{1}{ c|}{0.40/0.88} & \multirow{4}{*}{\begin{tabular}[c]{@{} c@{}}611.6\\ 1.8\end{tabular}} & \multicolumn{1}{c|}{0.28/0.65} & \multicolumn{1}{c|}{\multirow{4}{*}{\begin{tabular}[c]{@{}c@{}}39.5\\ 25.3\end{tabular}}} & \multicolumn{1}{c|}{0.27/0.63} & \multicolumn{1}{c|}{\multirow{4}{*}{\begin{tabular}[c]{@{}c@{}}36.7\\ 27.2\end{tabular}}} & \multicolumn{1}{c|}{0.25/0.65} & \multicolumn{1}{c|}{\multirow{4}{*}{\begin{tabular}[c]{@{}c@{}}26.3\\ 38.0\end{tabular}}} & \multicolumn{1}{c|}{0.28/0.65} & \multicolumn{1}{c|}{\multirow{4}{*}{\begin{tabular}[c]{@{}c@{}}14.3\\ 69.9\end{tabular}}} & \multicolumn{1}{c|}{\textbf{0.39/0.94}} & \multicolumn{2}{c}{\multirow{4}{*}{\begin{tabular}[c]{@{}c@{}}\textbf{11.5}\\ \textbf{87.0}\end{tabular}}} \\
\multicolumn{1}{c|}{} & \multicolumn{2}{l|}{BQMall} & \multicolumn{1}{ c|}{0.51/1.03} & \multicolumn{1}{c|}{} & \multicolumn{1}{ c|}{0.74/1.44} &  & \multicolumn{1}{c|}{0.33/0.88} & \multicolumn{1}{c|}{} & \multicolumn{1}{c|}{0.33/0.80} & \multicolumn{1}{c|}{} & \multicolumn{1}{c|}{0.28/0.68} & \multicolumn{1}{c|}{} & \multicolumn{1}{c|}{0.34/0.88} & \multicolumn{1}{c|}{} & \multicolumn{1}{c|}{\textbf{0.50/1.25}} & \multicolumn{2}{c}{} \\
\multicolumn{1}{c|}{} & \multicolumn{2}{l|}{PartyScene} & \multicolumn{1}{ c|}{0.22/0.73} & \multicolumn{1}{c|}{} & \multicolumn{1}{ c|}{0.51/1.46} &  & \multicolumn{1}{c|}{0.13/0.45} & \multicolumn{1}{c|}{} & \multicolumn{1}{c|}{0.17/0.58} & \multicolumn{1}{c|}{} & \multicolumn{1}{c|}{0.13/0.48} & \multicolumn{1}{c|}{} & \multicolumn{1}{c|}{0.16/0.48} & \multicolumn{1}{c|}{} & \multicolumn{1}{c|}{\textbf{0.34/1.15}} & \multicolumn{2}{c}{} \\
\multicolumn{1}{c|}{} & \multicolumn{2}{l|}{BasketballDrill} & \multicolumn{1}{ c|}{0.48/0.90} & \multicolumn{1}{c|}{} & \multicolumn{1}{ c|}{0.66/1.27} &  & \multicolumn{1}{c|}{0.38/0.88} & \multicolumn{1}{c|}{} & \multicolumn{1}{c|}{0.35/0.68} & \multicolumn{1}{c|}{} & \multicolumn{1}{c|}{0.33/0.68} & \multicolumn{1}{c|}{} & \multicolumn{1}{c|}{0.39/0.78} & \multicolumn{1}{c|}{} & \multicolumn{1}{c|}{\textbf{0.57/1.37}} & \multicolumn{2}{c}{} \\ \hline
\multicolumn{1}{c|}{\multirow{4}{*}{D}} & \multicolumn{2}{l|}{RaceHorses} & \multicolumn{1}{ c|}{0.51/1.13} & \multicolumn{1}{c|}{\multirow{4}{*}{\begin{tabular}[c]{@{} c@{}}164.3\\ 32.2\end{tabular}}} & \multicolumn{1}{ c|}{0.60/1.44} & \multirow{4}{*}{\begin{tabular}[c]{@{} c@{}}203.7\\ 7.3\end{tabular}} & \multicolumn{1}{c|}{0.33/0.83} & \multicolumn{1}{c|}{\multirow{4}{*}{\begin{tabular}[c]{@{}c@{}}24.3\\ 41.2\end{tabular}}} & \multicolumn{1}{c|}{0.32/0.75} & \multicolumn{1}{c|}{\multirow{4}{*}{\begin{tabular}[c]{@{}c@{}}12.1\\ 82.6\end{tabular}}} & \multicolumn{1}{c|}{0.31/0.73} & \multicolumn{1}{c|}{\multirow{4}{*}{\begin{tabular}[c]{@{}c@{}}7.0\\ 142.9\end{tabular}}} & \multicolumn{1}{c|}{0.38/0.83} & \multicolumn{1}{c|}{\multirow{4}{*}{\begin{tabular}[c]{@{}c@{}}\textbf{3.5}\\ \textbf{285.7}\end{tabular}}} & \multicolumn{1}{c|}{\textbf{0.44/1.08}} & \multicolumn{2}{c}{\multirow{4}{*}{\begin{tabular}[c]{@{}c@{}}9.1\\ 109.9\end{tabular}}} \\
\multicolumn{1}{c|}{} & \multicolumn{2}{l|}{BQSquare} & \multicolumn{1}{ c|}{-0.01/0.15} & \multicolumn{1}{c|}{} & \multicolumn{1}{ c|}{0.79/1.14} &  & \multicolumn{1}{c|}{0.09/0.25} & \multicolumn{1}{c|}{} & \multicolumn{1}{c|}{0.20/0.38} & \multicolumn{1}{c|}{} & \multicolumn{1}{c|}{0.13/0.18} & \multicolumn{1}{c|}{} & \multicolumn{1}{c|}{0.20/0.38} & \multicolumn{1}{c|}{} & \multicolumn{1}{c|}{\textbf{0.37/0.75}} & \multicolumn{2}{c}{} \\
\multicolumn{1}{c|}{} & \multicolumn{2}{l|}{BlowingBubbles} & \multicolumn{1}{ c|}{0.39/1.20} & \multicolumn{1}{c|}{} & \multicolumn{1}{ c|}{0.62/1.95} &  & \multicolumn{1}{c|}{0.21/0.68} & \multicolumn{1}{c|}{} & \multicolumn{1}{c|}{0.23/0.68} & \multicolumn{1}{c|}{} & \multicolumn{1}{c|}{0.18/0.58} & \multicolumn{1}{c|}{} & \multicolumn{1}{c|}{0.22/0.65} & \multicolumn{1}{c|}{} & \multicolumn{1}{c|}{\textbf{0.37/1.23}} & \multicolumn{2}{c}{} \\
\multicolumn{1}{c|}{} & \multicolumn{2}{l|}{BaskerballPass} & \multicolumn{1}{ c|}{0.63/1.38} & \multicolumn{1}{c|}{} & \multicolumn{1}{ c|}{0.85/1.75} &  & \multicolumn{1}{c|}{0.34/0.85} & \multicolumn{1}{c|}{} & \multicolumn{1}{c|}{0.34/0.78} & \multicolumn{1}{c|}{} & \multicolumn{1}{c|}{0.31/0.75} & \multicolumn{1}{c|}{} & \multicolumn{1}{c|}{0.35/0.85} & \multicolumn{1}{c|}{} & \multicolumn{1}{c|}{\textbf{0.48/1.28}} & \multicolumn{2}{c}{} \\ \hline
\multicolumn{1}{c|}{\multirow{3}{*}{E}} & \multicolumn{2}{l|}{FourPeople} & \multicolumn{1}{ c|}{0.66/0.85} & \multicolumn{1}{c|}{\multirow{3}{*}{\begin{tabular}[c]{@{} c@{}}345.4\\ 4.7\end{tabular}}} & \multicolumn{1}{ c|}{0.95/1.12} & \multirow{3}{*}{\begin{tabular}[c]{@{} c@{}}1.3K\\ 0.8\end{tabular}} & \multicolumn{1}{c|}{0.45/0.70} & \multicolumn{1}{c|}{\multirow{3}{*}{\begin{tabular}[c]{@{}c@{}}87.4\\ 11.4\end{tabular}}} & \multicolumn{1}{c|}{0.46/0.70} & \multicolumn{1}{c|}{\multirow{3}{*}{\begin{tabular}[c]{@{}c@{}}81.6\\ 12.3\end{tabular}}} & \multicolumn{1}{c|}{0.39/0.60} & \multicolumn{1}{c|}{\multirow{3}{*}{\begin{tabular}[c]{@{}c@{}}64.8\\ 15.4\end{tabular}}} & \multicolumn{1}{c|}{0.51/0.78} & \multicolumn{1}{c|}{\multirow{3}{*}{\begin{tabular}[c]{@{}c@{}}34.3\\ 29.2\end{tabular}}} & \multicolumn{1}{c|}{\textbf{0.69/1.02}} & \multicolumn{2}{c}{\multirow{3}{*}{\begin{tabular}[c]{@{}c@{}}\textbf{24.7}\\ \textbf{40.5}\end{tabular}}} \\
\multicolumn{1}{c|}{} & \multicolumn{2}{l|}{Johnny} & \multicolumn{1}{ c|}{0.55/0.55} & \multicolumn{1}{c|}{} & \multicolumn{1}{ c|}{0.75/0.85} &  & \multicolumn{1}{c|}{0.40/0.60} & \multicolumn{1}{c|}{} & \multicolumn{1}{c|}{0.38/0.40} & \multicolumn{1}{c|}{} & \multicolumn{1}{c|}{0.32/0.40} & \multicolumn{1}{c|}{} & \multicolumn{1}{c|}{0.41/0.50} & \multicolumn{1}{c|}{} & \multicolumn{1}{c|}{\textbf{0.93/0.64}} & \multicolumn{2}{c}{} \\
\multicolumn{1}{c|}{} & \multicolumn{2}{l|}{KristenAndSara} & \multicolumn{1}{ c|}{0.66/0.75} & \multicolumn{1}{c|}{} & \multicolumn{1}{ c|}{0.93/0.91} &  & \multicolumn{1}{c|}{0.49/0.68} & \multicolumn{1}{c|}{} & \multicolumn{1}{c|}{0.48/0.60} & \multicolumn{1}{c|}{} & \multicolumn{1}{c|}{0.42/0.60} & \multicolumn{1}{c|}{} & \multicolumn{1}{c|}{0.52/0.70} & \multicolumn{1}{c|}{} & \multicolumn{1}{c|}{\textbf{0.74/0.84}} & \multicolumn{2}{c}{} \\ \hline
\multicolumn{3}{c|}{Average} & \multicolumn{1}{ c|}{0.46/0.88} & \multicolumn{1}{ c|}{-} & \multicolumn{1}{ c|}{0.69/1.25} & \multicolumn{1}{ c|}{-} & \multicolumn{1}{c|}{0.30/0.66} & \multicolumn{1}{c|}{-} & \multicolumn{1}{c|}{0.30/0.63} & \multicolumn{1}{c|}{-} & \multicolumn{1}{c|}{0.26/0.58} & \multicolumn{1}{c|}{-} & \multicolumn{1}{c|}{0.32/0.67} & \multicolumn{1}{c|}{-} & \multicolumn{1}{c|}{\textbf{0.49/1.03}} & \multicolumn{2}{c}{\textbf{-}} \\ \hline
\end{tabular}%
}
\end{table*}

\subsection{Implementation Details}
Our proposed network is implemented based on PyTorch framework and MMEditing toolbox~\cite{mmediting2022}. During the training phase, we set batch size to 8 with 180$\times$180 clips from the raw videos and the corresponding compressed videos as training samples. The random flip and rotation augmentation are utilized to fully exploit the information in the samples. The Adam optimizer~\cite{kingma2014adam} and Cosine Annealing scheme~\cite{loshchilov2016sgdr} are utilized for optimizing. The total iterations numbers and initial learning rate of the first and second stages is set to 200K, $1\times10^{-4}$ and 100K, $1\times10^{-6}$. Our network is trained on 8 Telsa V100 GPUs.

During the inference phase, ST-LUTs will replace the corresponding convolutional layers. 
To evaluate the model's performance in an online setting, we conduct inference on the test set frame-by-frame. Following the same settings as ~\cite{guan2019mfqe}, we report quality enhancement on Y-channel in YUV/YCbCr space. The Peak Singal-to-Noise Ratio($\Delta$PSNR) and Structural Similarity($\Delta$SSIM) will be used as metrics to judge the model's enhancement capability.
We further evaluate the latency(LT) and frame per second(FPS) of the models which is crucial to the online-VQE:  
\begin{equation}\label{LT}
\begin{aligned}
LT = T_{process} + T_{wait}, \quad T_{wait} = N_{f} / FPS, 
\end{aligned}
\end{equation}
where $T_{process}$ refers to the single-frame inference time and $T_{wait}$ refers to the future frames waiting time required for some video based methods, $N_f$ represent the number of future frames used. Note that the LT metric ignores the frame reception and decoding time, which is fair for different methods, and we set $FPS$=30 in Eq. \ref{LT} as the general online video standard to facilitate comparison. 

\begin{table}[t]
\centering
\caption{Latency on 720P video and averaged $\Delta$PSNR/$\Delta$SSIM comparison on the test sequences of MFQE 2.0 Dataset at different QP values. FV-Required means the full compressed video is required for enhancement, thus latency cannot be calculated.   }\label{tab:Different_QP}
\resizebox{\textwidth}{!}{%
\begin{tabular}{c|c|c|c|c|ccccc}
\hline
Method     & Type   & LT(ms)      & Storage & GFLOPs & QP37      & QP42      & QP32      & QP27      & QP22      \\ \hline
\rowcolor{mygray} MFQEv1~\cite{yang2018multi}     & Bi-MF  & 345.4       & 1788K          & -           & 0.46/0.88 & 0.44/1.30 & 0.43/0.58 & 0.40/0.34 & 0.31/0.19 \\
\rowcolor{mygray} MFQEv2~\cite{guan2019mfqe}     & Bi-MF  & 320.1       & 255K           & 788       & 0.56/1.09 & 0.59/1.65 & 0.52/0.68 & 0.49/0.42 & 0.46/0.27 \\
\rowcolor{mygray} STDF~\cite{deng2020spatio}       & Bi-MF  & 85.6        & 487K           & 293       & 0.65/1.18 & -/-       & 0.64/0.77 & 0.59/0.47 & 0.51/0.27 \\
\rowcolor{mygray} PSTQE~\cite{ding2021patch}      & Bi-MF  & 1249.8      & 218K             & 679       & 0.69/1.25 & 0.69/1.86 & 0.67/0.83 & 0.63/0.52 & 0.55/0.29 \\
\rowcolor{mygray} RFDA~\cite{zhao2021recursive}       & Bi-MF  & 312.8       & 1250K          &  -          & 0.91/1.62 & 0.82/2.20 & 0.87/1.07 & 0.82/0.68 & 0.76/0.42 \\
\rowcolor{mygray} BasicVSR++~\cite{chan2022basicvsr++} & Bi-MF  & FV-Required & 7300K          & 423      & 0.95/1.80 & -/-       & 0.89/1.25 & -/-       & -/-       \\ \hline
\rowcolor{mygray} RNAN~\cite{zhang2019residual}       & SF     & 4329.4      & 8960K          & 6979      & 0.44/0.95 & -/-       & 0.41/0.62 & -/-       & -/-       \\ 
ARCNN~\cite{dong2015compression}      & SF     & 25.1        & 118K           & 108       & 0.23/0.45 & 0.29/0.96 & 0.18/0.19 & 0.18/0.14 & 0.14/0.08 \\
DnCNN~\cite{zhang2017beyond}      & SF     & 64.8        & 558K           & 516       & 0.26/0.58 & 0.22/0.77 & 0.26/0.35 & 0.27/0.24 & 0.29/0.18 \\
DS-CNN~\cite{yang2018enhancing}     & SF     & 81.6        & 948K           & 437       & 0.30/0.63 & 0.31/1.01 & 0.27/0.38 & 0.27/0.23 & 0.25/0.15 \\
Li et al.~\cite{li2017efficient}  & SF     & 87.4        & 464K           & 434           & 0.30/0.66 & 0.32/1.05 & 0.28/0.37 & 0.30/0.28 & 0.30/0.19 \\
DCAD~\cite{wang2017novel}       & SF     & 34.3        & 299K           & 275       & 0.32/0.67 & 0.32/1.09 & 0.32/0.44 & 0.32/0.30 & 0.31/0.19 \\ \hline
STLVQE     & Uni-MF & \textbf{24.7}        & 398K+228M      & \textbf{98}       & \textbf{0.49/1.03} & \textbf{0.50/1.60} & \textbf{0.45/0.64} & \textbf{0.41/0.40} & \textbf{0.34/0.24} \\ \hline
\end{tabular}
}
\end{table}

\subsection{Comparison with State of the art Methods}

In this subsection, we present the results of our extensive experiments on various QP compressed videos in Tab.~\ref{tab:QP37} and Tab.~\ref{tab:Different_QP}. Our proposed STLVQE network is compared with some widely used or state-of-art single-frame methods~\cite{dong2015compression, zhang2017beyond, wang2017novel, yang2018enhancing, li2017efficient, zhang2019residual}, as well as some efficient multi-frame methods~\cite{yang2018multi, guan2019mfqe, deng2020spatio, ding2021patch}. 
In online environment where video is streamed in real-time, Bi-direction Multi-Frame (Bi-MF) method necessitates waiting for future frames that have not yet been received in current time. 
This waiting period introduces uncontrollable latency to the stream. 
Consequently, as outlined in previous online works~\cite{yin2024online, lin2019tsm, maggioni2021efficient}, the Bi-MF method can not be directly applied to online scenarios and is predominantly utilized for offline processing. As a result, both Bi-MF and slow single frame methods(\textgreater 100ms) are placed in the first part in Tab.~\ref{tab:QP37} and Tab.~\ref{tab:Different_QP} for performance comparison, and demonstrate our favorable trade-off between performance and inference speed. 
All experiments were conducted on an NVIDIA RTX 3090 using a frame-to-frame style.

\textbf{Latency and FPS.} Tab. ~\ref{tab:QP37} presents the latency and FPS metrics for each VQE approach. We note that while low-latency VQE methods can process each frame at a lower latency and reduce video delay, they often struggle to handle high-resolution videos in real-time, which can lead to stuttering and reduced smoothness. Among the methods listed, only STLVQE can achieve FPS above 40 on 720P resolution videos, making it suitable for most online scenarios. Compared to the second fastest method DCAD, STLVQE's processing time on 720p video is 9.6 milliseconds faster and outperforms it by a considerable margin, improving PSNR and SSIM metrics by 0.17dB and $0.36\times10^{-2}$, respectively.

Tab.~\ref{tab:Different_QP} shows the latency of some large-scale high-latency temporal VQE methods. In contrast, high-latency VQE methods require waiting for future frames or taking a long time for inference, which can cause uncontrollable video delay. The latency of the two high-latency methods in Tab.~\ref{tab:QP37} we evaluated exceeds 400ms, much higher than the latency caused by STLVQE. Moreover, when considering only the FPS metric, STLVQE achieves a significantly faster speed than the single-frame VQE method RNAN(FPS = 0.2) by a factor of 175 on 720P video. Compared with MFQE 1.0, our method is 8.6 times faster with a better enhancement performance. Additionally, STLVQE has a more significant speed advantage on higher resolution videos such as 1080P and 2K. 

\begin{table}[t]
\centering
\small
\caption{Ablation study of Temporal Cache(TC), Reference Frame Window(RFW), Deform Operation(DF), and Preliminary Complement Residual(PC). The latter three sections share the features extracted by MAFE.}\label{tab:Ablation_Structure}
\begin{tabular}{cccc|ccc}
\hline 
RFW   &\textbf{TC}    &\textbf{DF}   &\textbf{PC}   &$\Delta$P/$\Delta$S &Time &Params(K) \\
\hline
 $\checkmark$ &$\checkmark$  &  &   &0.06/0.01  &10.85 &$<$1 \\
$\checkmark$  &$\checkmark$  &$\checkmark$  &   &0.31/0.51  &21.85    &352 \\
$\checkmark$  &$\checkmark$  &  &$\checkmark$     &0.39/0.87  &20.12  &149   \\
$\checkmark$  &  &$\checkmark$  &$\checkmark$   &0.49/1.03  &45.61   &398 \\
  &$\checkmark$  &$\checkmark$  &$\checkmark$     &0.46/0.98  &24.32  &398   \\  
$\checkmark$  &$\checkmark$  &$\checkmark$  &$\checkmark$     &0.49/1.03  &24.65  &398   \\
 
\hline
\end{tabular}
\end{table}


\textbf{Enhancement performance.} 
In Tab.~\ref{tab:QP37} and Tab.~\ref{tab:Different_QP}, we present the performance of STLVQE compared to other commonly used or sota VQE methods for each video at QP=37 and for different QP compressed videos, respectively. Our method outperforms all the listed low-latency methods. 
In comparison with the high-latency VQE method, STLVQE reaches higher level with RNAN at QP=37 and QP=32, with only 2.6\% and 0.6\% of RNAN's number of parameters and latency time (720P), respectively.
The $\Delta$SSIM of STLVQE is higher than that of the MFQE method for all five QPs, indicating that our method produces an enhanced image with a more stable image structure. The specific visualization results and analysis can be found in \textit{Qualitative Comparison} section. STDF is a robust VQE method, but we believe that a stable latency and higher frame rate are more important for enhancing the user's viewing experience.

\textbf{Memory Size \& GPU Consumption.} As shown in Tab.~\ref{tab:Different_QP}, STLVQE 
has space occupation of 398KB for the model and 228MB for the LUT(utilizing less than 5\% memory of a low end GPU like GTX 1060), achieving real-time processing capability. Since most deep networks are replaced with ST-LUTs, inferencing STLVQE only needs 98 GFLOPs on  a 720P frame. Note that GPU computing unit are more expensive than the memory. 

\subsection{Ablation Study}

To validate the effectiveness of our proposed new VQE framework, we perform ablation experiments and specific analysis on three components. The results of the runtime per frame is computed on 720P videos at QP = 37. 

\textbf{Effectiveness of the propagation module.} The results are shown in the 5th and 6th row of Tab. ~\ref{tab:Ablation_Structure}. We found that selecting frames based on QP values in RFW can bring a 0.03dB $\Delta$PSNR performance improvement compared to randomly selecting frames in the RFW, without adding parameters to the network and only increasing the inference time by 0.33 ms for comparing QP values of each frame. 
Furthermore, as shown in rows 4 and 6 of Tab.~\ref{tab:Ablation_Structure}, the use of the TC structure can significantly reduce the inference time by 21.0 ms, which is consumed by the repeated computation of the reference frame information. We achieve this reduction in time by merely occupying the GPU memory with information of several frames in the RFW. 

\textbf{Effectiveness of the alignment module.} We mainly validate the effect of the deform operation of the reference fram in the Alignment module. As shown in the first row of Tab.~\ref{tab:Ablation_Structure}, when only using the ST-LUTs structure and discard the whole alignment module and PC operation, the result only has a 0.06 dB PSNR improvement compared to HEVC video.
When the Deform operation is applied, the network can capture more accurate temporal information, and the $\Delta$PSNR reaches 0.31 dB. Due to the lightweight convolutional layers, the alignment module only adds 11.6 ms of inference time to the network. 
\begin{table}[t]
\centering
\small
\caption{Ablation study of STLUTs and the CNN enhancement network. The runtime is computed on a 720P video.}\label{tab:Ablation_LUT}
\begin{tabular}{c|c|c|c}
\hline 
Enhancement Module   &$\Delta$P/$\Delta$S   &Time(ms)   &Params(K) \\
\hline
S$-$LUT&0.39/0.83  &23.52  &398    \\
S$-$LUT $+$ T$-$LUT$-$1 &0.45/1.02  &24.12  &398     \\
ST$-$LUTs  &0.47/1.02  &24.65  &398    \\
ST$-$LUTs (on $\Tilde{x}_i$)  &0.49/1.03  &24.65  &398    \\
CNN Network &0.50/1.08  &147.02  &442\\
 
\hline
\end{tabular}
\end{table}

\textbf{Effectiveness of the enhancement module.} 
As shown in row 3 of Tab.~\ref{tab:Ablation_Structure}, using PC without the Deform operation achieves a significant $\Delta$PSNR of 0.39 dB, demonstrating the effectiveness of PC in providing initial compensation for spatial information and extending the receptive field, ultimately enhancing the performance of ST-LUTs. Finally, after combining two parts, the network achieves 0.49 dB $\Delta$PSNR. PC alone contributes 0.18 dB to the network and adds only 2.16 ms of inference time.

Tab.~\ref{tab:Ablation_LUT} provides an analysis of the impact of each LUT in the enhancement module. In the first three rows of the table, we attempted to enhance the original image $x_i$ with the results of different LUT combinations. Results in the first and second rows show that using only the S-LUT with T-LUT-1 does not fully improve the network due to T-LUT-1 not utilizing all pixels within a kernel when learning the alignment offset. However, when T-LUT-2 is used for network training along with T-LUT-1, the reference frame can be better aligned, resulting in a $\Delta$PSNR improvement of 0.08 dB.
The 4th row shows the enhancing result on $\Tilde{x}_i$, reducing the learning difficulty of the ST-LUTs, which improving the performance by 0.02 dB. Compared to the CNN layers replaced by LUTs, ST-LUTs sacrifice only 0.01 dB $\Delta$PSNR performance while reducing the number of parameters by 44K and the inference time by 122.37 ms.

The above ablation studies have sufficiently demonstrated  that the MAFE structure of shared features in STLVQE with the redesigned three new modules can sufficiently reduce the computational redundancy of the network and obtain satisfactory inference speed-performance trade-offs.

\subsection{Qualitative Comparison}\label{Visual}

\begin{figure*}[t]
  \centering
  \includegraphics[width=\linewidth,trim=20 85 70 40,clip]{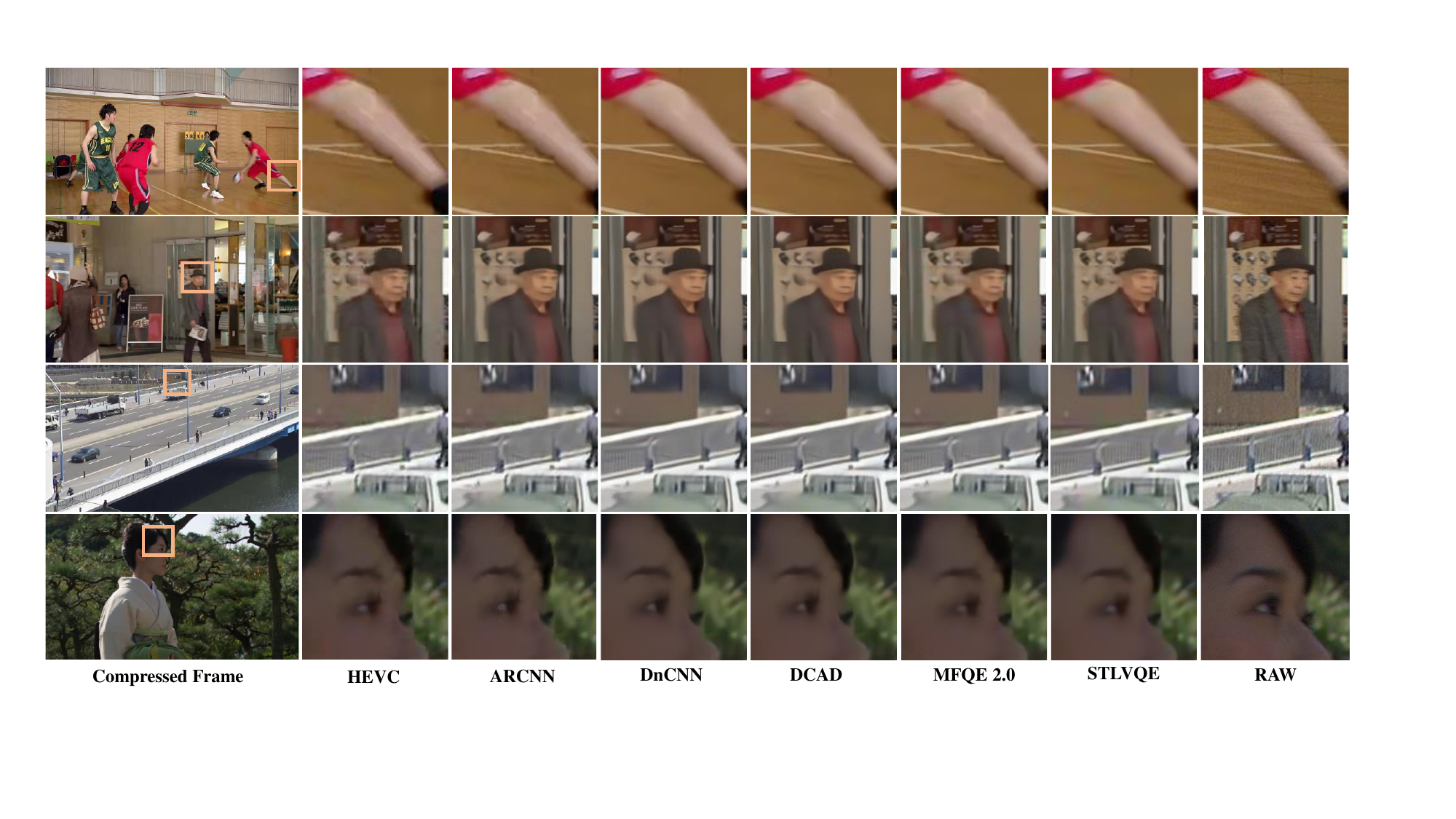}
  \caption{Qualitative results at QP 37. 
  The STLVQE method successfully enhances the left leg of the athlete, the face of the old man, the railings of the bridge and the eyebrow of the woman in each frame, respectively.   }\label{Quality}
\end{figure*}

Fig.~\ref{Quality} provides 4 examples of the qualitative results of our network on compressed video at \textit{QP=37}. The quality of these images deteriorates significantly in compressed scenes, showing a variety of image defects (e.g. blurring legs, missing bridge railings). With the help of temporal information, STLVQE achieves better visual results than single-frame methods with significantly reduced runtime. Compared to large-scale multi-frame methods like MFQE 2.0, STLVQE can achieve competitive capabilities in detail and structural restoration.

\section{Conclusion}
In this paper, we propose a lightweight and rapid method called STLVQE based on the constraints of online video quality enhancement tasks. STLVQE greatly reduce the computation redundancy and uses a Spatial-Temporal Look-up Table structure to extract spatio-temporal information from the video with minimal time consumption. The extensive experiments prove that our proposed method can outperform most low-latency VQE methods, achieve competitive performance with high-latency VQE methods which cannot be directly utilized on Online-VQE tasks. Furthermore, STLVQE achieves real-time processing of 720P resolution videos, demonstrating a good trade-off between enhancement performance and inference speed. 
\clearpage  

\section*{Acknowledgements}
This work was supported by National Natural Science Fund of China (62076184, 61976158, 61976160, 62076182, 62276190), in part by Fundamental Research Funds for the Central Universities and State Key Laboratory of Integrated Services Networks (Xidian University), in part by Shanghai Innovation Action Project of Science and Technology (20511100700) and Shanghai Natural Science Foundation (22ZR1466700).

%
%
\bibliographystyle{splncs04}
\bibliography{main}
\end{document}


\title{Supplementary Material of Online Video Quality Enhancement with Spatial-Temporal Look-up Tables} 

\titlerunning{Abbreviated paper title}

\author{First Author\inst{1}\orcidlink{0000-1111-2222-3333} \and
Second Author\inst{2,3}\orcidlink{1111-2222-3333-4444} \and
Third Author\inst{3}\orcidlink{2222--3333-4444-5555}}

\authorrunning{F.~Author et al.}

\institute{Princeton University, Princeton NJ 08544, USA \and
Springer Heidelberg, Tiergartenstr.~17, 69121 Heidelberg, Germany
\email{lncs@springer.com}\\
\url{http://www.springer.com/gp/computer-science/lncs} \and
ABC Institute, Rupert-Karls-University Heidelberg, Heidelberg, Germany\\
\email{\{abc,lncs\}@uni-heidelberg.de}}

\maketitle

\begin{figure}[t]
  \centering
  \includegraphics[width=\linewidth,trim=0 125 0 60,clip]{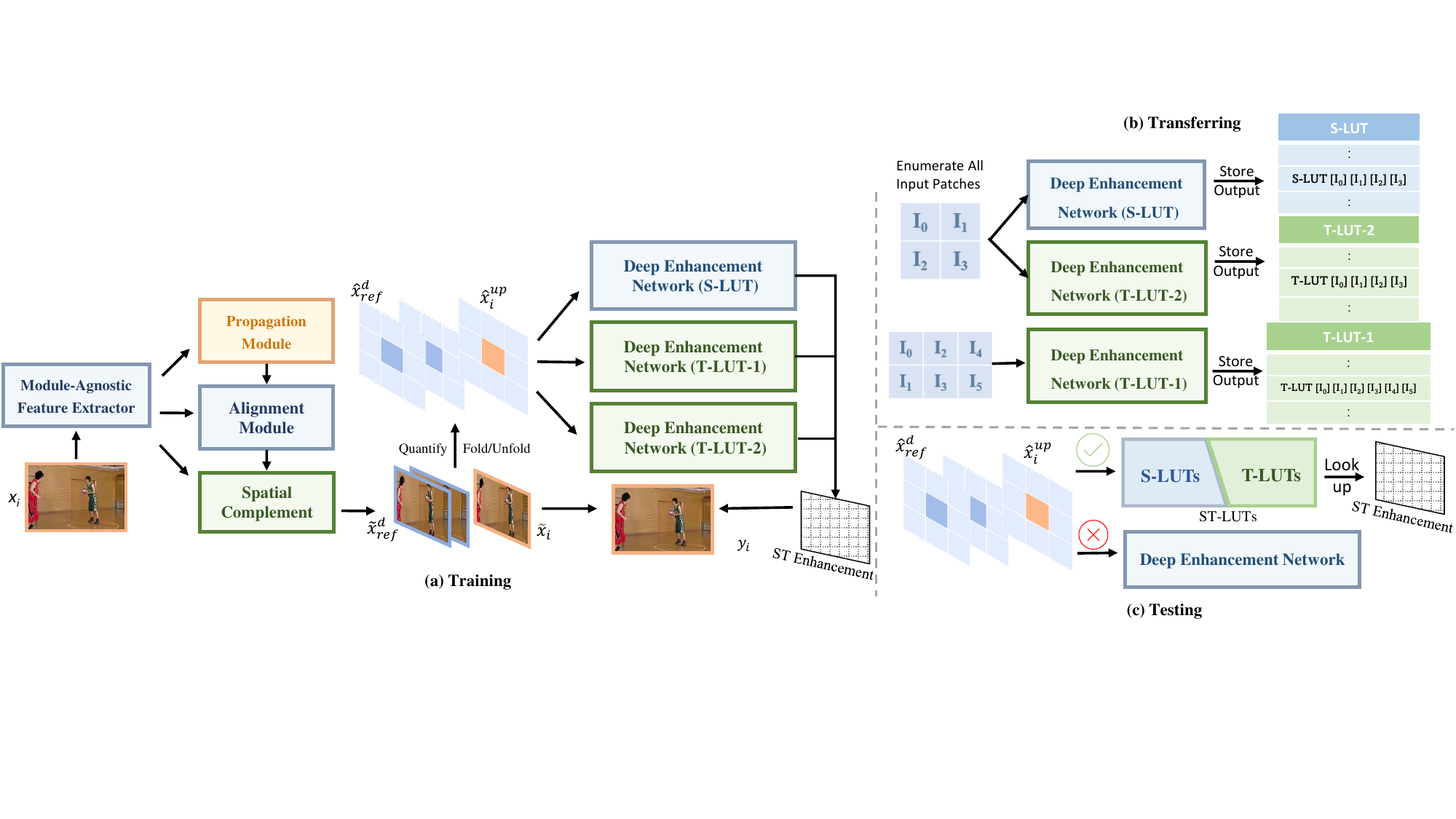}
  \caption{Details of whole procedure before the inference phase of STLVQE.}\label{Procedure}
\end{figure}

\section{Network Architecture}
In this section we will illustrate the architecture of our propagation module and the deep network in the alignment and enhancement module in the training phase.


\textbf{The procedure of the STLVQE before the inference phase. }
The network flow of STLVQE is detailed illustrated in the Fig.~\ref{Procedure} and can be summarized as following steps: 
(1)\textbf{Propagation:} Selects the reference frames by the QP value and delivers the information to the alignment module, meanwhile updates the Temporal Cache.
(2)\textbf{Alignment:} Compute offset based on features, then perform deformation on $\Tilde{x}_{ref}$ to align the reference frames with the current frame. 
(3)\textbf{Enhancement:} 
Spatial compensation is firstly applied to the current frame, which can fully utilize the sharing feature and increase the receptive field of ST-LUTs.
Then $\Tilde{x}_{ref}^{d}$ and $\Tilde{x}_{i}$ are quantized and input into Deep LUT Network(DLN) for the Spatial-Temporal Enhancement residual result. 
After training, we enumerate all the inputs and store the corresponding DLN results into ST-LUTs(Fig.~\ref{Procedure}(b)), these transferred ST-LUTs will replace the DLN during inference phase to reduce the inference time and computing resource consumption(Fig.~\ref{Procedure}(c)). The only difference between training and inference procedure is the replacement of the Deep Network with the transferred ST-LUTs. The digit in Fig.4 represents the pixels position in each kernel which is used for DLN training and LUT retrieval. 

\textbf{The propagation module. }
The architecture of the propagation module is illustrated in Fig. \ref{window}. We explain this module in more detail in this section.

The QP value indicates the degree of compression applied to a particular frame during the encoding process, with a smaller QP value indicating less loss and higher image quality. By using the QP value, we can select fewer but higher quality frames among multiple candidate frames without introducing extra computation. 
Specifically, we set up a Reference Frame Window (RFW) during the inference phase, as shown in Fig.~\ref{window}, which contains $W$ frames $\{x_{i-W}.... . x_{i-1}\}$. Then, we select two frames with the smallest QP values as $x_{ref}$ from this window. By leveraging the prior knowledge of compressed VQE tasks and directly selecting reference frames based on their QP values, we can select high-quality frames without the need for additional network structures or increasing computational costs.


The propagation module of many previous methods require an additional feature extraction step for the reference frames, resulting in increased computational costs that can hinder the feasibility of online VQE tasks.
To mitigate this issue, we introduce a cache-like structure known as the Temporal Cache (TC). The TC stores reusable features of past frames that were already extracted by the alignment module in previous process steps, thereby eliminating the need to re-extract features for each subsequent frame. 

\begin{figure}[h]
  \centering
  \includegraphics[width=0.7\linewidth,trim=100 125 160 30,clip]{fig/Reference Frame Windwo.pdf}
  \caption{Details of Reference Frame Window and Temporal Cache mechanism in the propagation module.}\label{window}
\end{figure}

Specifically, when enhancing the i-th frame $x_i$, STLVQE caches the intermediate feature map $Feat_i$ extracted by the alignment module, as well as the enhanced frame, in the information cache (TC).  
During reference frame selection, the features and enhanced result of $x_{ref1}$ and $x_{ref2}$ are directly retrieved from TC without extra feature extraction process. 

\begin{figure*}[t]
  \centering
  \includegraphics[width=\linewidth,trim=10 80 20 50,clip]{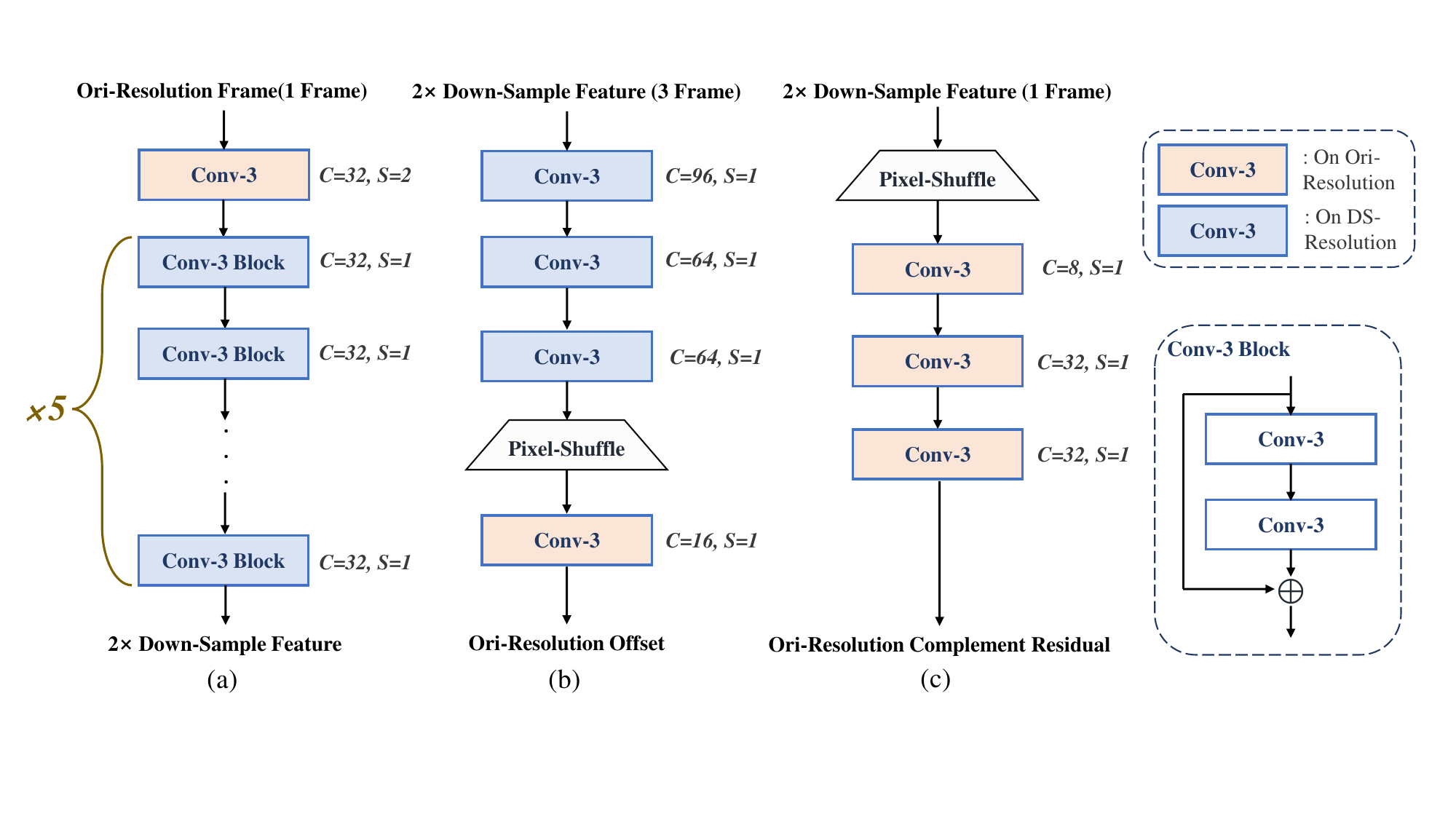}
  \caption{The architecture of each part in the STLVQE: (a) The feature extraction module. (b) The offset prediction module. (c) The spatial complement module. C,S denote the channel and stride of the each convolution layer respectively.}\label{alignment_archit}
\end{figure*}

\textbf{Module-Agnostic Feature Extraction module. }
The architecture of MAFE is illustrated in (a) in Fig.~\ref{alignment_archit}. To enable the extraction of robust module-sharing features, this part uses an 11-layer 3$\times$3 convolution layers, which is computed on a downsampled 2x image, so MAFE does not excessively increase the inference time of the network.


\textbf{The alignment module and spatial complement.} As illustrated in (b) and (c) in Fig.~\ref{alignment_archit}, the alignment module and the spatial complement network only contain 4 and 3 convolution layers and two pixel-shuffle layers (the activation function is omitted here). To reduce the network's inference time, we upsample it back to the original resolution in the last layer of the offset prediction network and the last two computation layers of the spatial complement network. 

\begin{figure}[t]
  \centering
  \includegraphics[width=\linewidth,trim=150 80 120 50,clip]{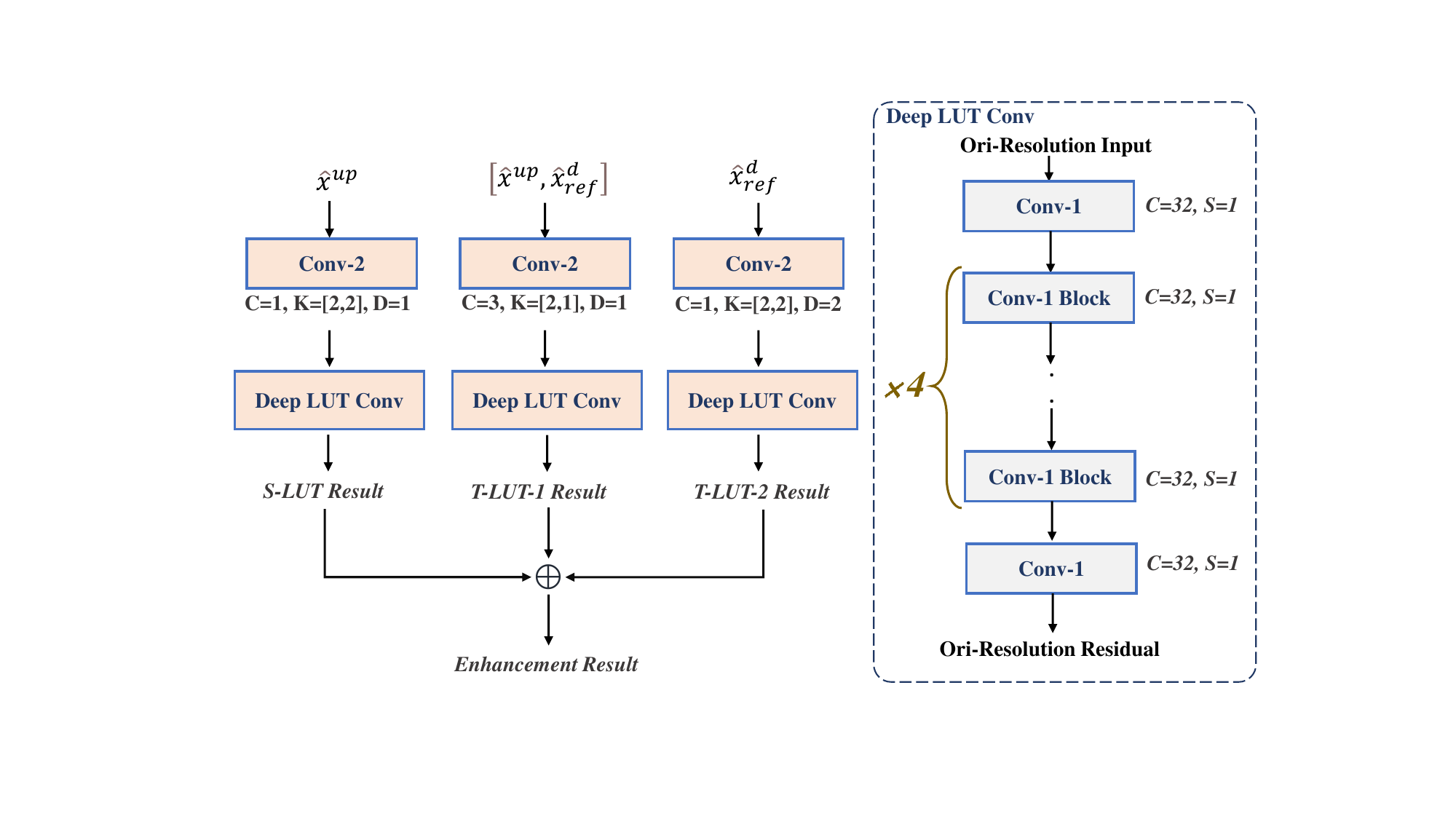}
  \caption{The architecture of three deep network in the enhancement module, these deep networks will be replaced by S-LUT, T-LUT-1 and T-LUT-2 in the inference stage. C,K,D denote the channel, kernel size and dilation value of the first convolution respectively.}\label{enhancement_archit}
\end{figure}

\textbf{The deep network replaced by ST-LUTs.} During the training stage, STLVQE's enhancement module uses three parallel deep networks. However, in the inference stage, these networks are replaced with S-LUT, T-LUT-1, and T-LUT-2, respectively. Fig.~\ref{enhancement_archit} illustrates that these networks share the same structure, except for the first convolution layer. In addition to the first layer, the subsequent structure of each network comprises 10 layers of 1*1 convolution.

The first convolution layer of each depth network determines the selection of LUT index pixels. As shown in Fig. 3 in the main text, the S-LUT network selects four neighboring pixels as indexes with the kernel size of 2. The T-LUT-1 network selects two pixel values as reference in both of the $\hat{x}_i^{up}$ and $\hat{x}^{d}_{ref}$, at which time these three frames are concatenated together on the channel dimension, and a convolution layer with channel=3 and kernel size=[2,1] is used to extract the temporal information. Finally, for T-LUT-2, we extract the 4 edge pixels in each 3×3 kennel in $\hat{x}^{d}_{ref}$. A convolution layer with kernel size of 2 and dilation of 2 is utilized for dilated convolution. Additionally, input frames are subjected to a rotation ensemble operation to further expand the receptive field of the LUTs.

\section{Details of the LUT Storage and Query}

After training the enhancement module deep network, the input and output mappings of the network with LUT can be stored. In our method, the output result of the enhancement module is an enhancing residual map of $\Tilde{x}_i$, which consists of float-type data. Unlike previous methods~\cite{jo2021practical, ma2022learning} that use LUTs for super-resolution(SR) tasks, our input-output pixel ratio is 4:1, i.e., four pixels are used to index the enhanced residual value of one pixel. This significantly reduces the storage space required for LUTs compared to the 4:16 ratio used in 4$\times$ upscaling SR tasks. Therefore, we set the sampling interval size to 4 in S-LUT and T-LUT-2 to minimize network performance loss due to sampling. As a result, the SLUT and T-LUT-2 tables in our approach has a size of $((2^8/2^2) + 1)^4\times 32bit = 68.09MB$, the T-LUT-1 tables has a size of $((2^8/2^4) + 1)^6\times 32bit = 92.08MB$. STLVQE stores a total of 3 LUT tables, consuming a total of $68.09 \times 2 + 92.08 = 228.26MB$.

We have adopted the tetrahedral interpolation method~\cite{kasson1995performing} for the interpolation operation, which requires only 5 multiplications with the values of the 5 bounding vertices of the 4-simplex geometry. This method is known to be significantly faster than other commonly used interpolation methods. Unlike the interpolation stage of SRLUT, our LUT input is split into the most significant bits (MSB) and least significant bits (LSB), with 6/2 and 4/4 bits allocated to each in S-LUT and T-LUT-1, respectively, due to our sampling interval of 4. In order to accelerate the interpolation process and reduce the time consumed by the if-else logic judgment, we have employed CUDA to accelerate the interpolation, based on the approach proposed in \cite{yin2023online}.




\section{Experiment}

\begin{table}[]
\caption{Comparison of STLVQE latency with other temporal VQE methods.}\label{latency}
\centering
\resizebox{0.7\columnwidth}{!}{%
\begin{tabular}{c|c|c}
\hline
Method & Parameter Size & Latency(On 720P) \\ \hline
MFQE 1.0\cite{yang2018multi} & 1788K & 345.4 ms \\
NL-ConvLSTM\cite{xu2019non} & - & \textgreater 345.4 ms \\
MFQE 2.0\cite{guan2019mfqe} & 255K & 320.1 ms \\
STDF-R3\cite{deng2020spatio} & 487K & 190.1 ms \\
RFDA\cite{zhao2021recursive} & 1250K & 312.8 ms \\
BasicVSR++\cite{chan2022basicvsr++} & 7300K & - (Full video required) \\ 
STLVQE & 398K & 24.7 ms \\ \hline
\end{tabular}%
}
\end{table}

\textbf{The latency comparison with temporal VQE method.} The Tab.~\ref{latency} lists the temporal VQE methods proposed in recent years, the latency of these methods is generally high and all of them use future frames as reference frames, using these methods on online videos will bring serious latency and lag phenomenon to the video. The STLVQE's per-frame latency on 720P video is only about 24 ms, which meets the needs of most online scenarios.

\begin{table}[]
\caption{Ablation studies on the number of the query pixels for the ST-LUTs structure.   }\label{samplingpixels}
\centering
\begin{tabular}{c|c|c|c}
\hline
RF & LUT & LUT Size(I=$2^2$) & LUT Size(I=$2^4$) \\ \hline
1 Pixel & 1D & 260 B  & 68 B\\
2 Pixels & 2D & 17 KB & 1 KB \\
3 Pixels & 3D & 1 MB & 19 KB \\
4 Pixels & 4D & 68 MB(S-LUT/T-LUT-2)  & 326 KB\\
5 Pixels & 5D & 4 GB & 5 MB \\ 
6 Pixels & 6D & 281 GB & 92 MB(T-LUT-1) \\ 
7 Pixels & 7D & 18 TB & 2 GB \\\hline
\end{tabular}
\end{table}

\textbf{The query pixels number of ST-LUTs.} As shown in Tab.~\ref{samplingpixels}, the LUT sizes increase exponetially as the number of pixels in the LUT input increases. We follow the approach~\cite{jo2021practical, li2022mulut, ma2022learning} proposed in the past, using a receptive size of $2^2$ for S-LUT and T-LUT-2. Looking at more pixels will portantially make T-LUTs expoit more temporal information, in practice it will also increase the number or the size of the look-up table and negatively affect our model's efficiency, so we set the receptive size as $3\times2$ for the T-LUT to balance the performance-speed trade-off. 

As an online-VQE method, it is cucial to maintain the efficiency of the algorithm, and the current configuration of the T-LUTs is a relatively good trade-off between performance and efficiency. 
Increasing the receptive field of the LUT is still a open problem and we will continue to explore better ways to obtain high-quality temporal information with LUT based method in the future works.


\begin{table}[]
\caption{The latency and performance comparisons on LDV 1.0 dataset.}\label{ldv1.0}
\centering
\begin{tabular}{c|c|c}
\hline
Method & Latency & PSNR \\ \hline
HEVC & - & 30.54dB \\
ARCNN & 24.1ms & 30.74dB \\
DnCNN & 64.8ms & 30.92dB \\
QECNN & \textgreater 64.8ms & 30.98dB \\
MFQE & 345.4ms & 31.12dB \\
STLVQE & 24.7ms & 31.21dB \\ \hline
\end{tabular}
\end{table}

\textbf{The results on LDV 1.0 dataset. }
We tested the PSNR enhancement of STLVQE on LDV 1.0~\cite{yang2021ntire} and displayed it in Tab.\ref{ldv1.0}, while we list the performance of HEVC video, ARCNN~\cite{dong2015compression}, DnCNN~\cite{zhang2017beyond}, QECNN~\cite{yang2018enhancing}, and MFQE~\cite{yang2018multi} methods on this dataset as comparison. This experiment once again verifies the conclusions we have reached in our main paper.

In our main paper, we following a standard evaluation protocol widely used by previous VQE works (e.g., MFQE 1.0~\cite{yang2018multi}, MFQE 2.0~\cite{guan2019mfqe}, STDF~\cite{deng2020spatio}, RFDA~\cite{zhao2021recursive}, etc.), where the proposed method is evaluated on the standard VQE benchmark named MFQE 2.0. Most of the SOTA methods only release their performance on MFQE 2.0, making it difficult of us to evaluate comparison methods on the other datasets in a relatively fair quantitative manner.

It is also worth mentioning that in the relatively new field of VQE, as the only well accepted benchmark dataset for now, the MFQE 2.0 dataset is carefully designed and covers wide range of diversified compression configurations and data distributions. Specifically, it consists of videos with 10 different resolutions and compressed with 5 different QP values. This is why most of the existing works believes MFQE 2.0 alone is enough to evaluate existing VQE methods.

\begin{table}[h]
\footnotesize
\centering
\caption{Hyper-parameter of deep enhancement network.}
\begin{tabular}{c|c|c}
\hline 
Block Number &Channel Number  &$\Delta$PSNR/$\Delta$SSIM \\
\hline
3 &32  &0.45/0.92  \\
4 &32  &0.49/1.03 (Ours) \\
4 &64  &0.49/1.03   \\
5 &32  &0.48/1.00  \\
 
\hline
\end{tabular}
\label{deepenhancenetwork}
\end{table}

\textbf{LUT Network architecture. } Tab.~\ref{deepenhancenetwork} shows how network complexity affects VQE performance, where network with different  block (depth) and channel (width) are compared. Because the network replaced by LUT is constrained by a small receptive field($\leq$6 pixels), a more complex network does not increase the performance boundlessly, even lead to some drop due to optimization difficulties. This conclusion is consistent with SR-LUT~\cite{jo2021practical}.

\begin{table}[h]
\renewcommand{\arraystretch}{0.3}
\footnotesize
\centering
\caption{T-LUT structure.}
\resizebox{0.8\textwidth}{!}{%
\begin{tabular}{c|c|ccccc}
\hline 
T-LUT-1 RF &LT(ms) &QP37  &QP42 &QP32 &QP27 &QP22 \\
\hline
2\times2(twice \, query) &25.1 &0.44/0.92 &0.46/1.42 &0.42/0.65  &0.38/0.37 &0.31/0.20   \\
2\times3(once \, query) &24.7 & 0.49/1.03 & 0.50/1.60 & 0.45/0.64 & 0.41/0.40 & 0.34/0.24 \\
 
\hline
\end{tabular}\label{tlut}
}

\vspace{-10pt}
\end{table}

\textbf{LUT Query Policy. } As shown Fig. 3, we already design three types of patch permutation strategies and the ablation study results are in Tab. 4 in our main paper.
 We further experiment another type of permutation strategy with  a 2$\times$2 receptive field LUT, and query it once per reference frame (2 pixels in $x$ and 2 pixels in $x_{ref}$), the ablation results are as follows in Tab~\ref{tlut}. Experiments demonstrate the importance of temporal receptive field size. There are still many LUT permutation options, which we will continue to explore in our future work.

\clearpage  

%
%
\bibliographystyle{splncs04}
\bibliography{main}